\documentclass[manuscript, screen]{jair}

\setcopyright{cc}
\copyrightyear{2026}

\RequirePackage[
  datamodel=acmdatamodel,
  style=acmauthoryear,
  backend=biber,
  giveninits=true,
  uniquename=init
  ]{biblatex}

\usepackage{amsmath} 
\usepackage{array}
\usepackage{multirow}
\usepackage{subcaption}
\usepackage{booktabs,threeparttable}

\usepackage{enumitem}

\usepackage{xcolor,soul}
\definecolor{cbgreen}{HTML}{44AA99}
\definecolor{cbblue}{HTML}{88CCEE}
\definecolor{cborange}{HTML}{DDCC77}

\definecolor{LineNumGray}{gray}{0.53}
\usepackage{algorithm}
\usepackage{algorithmicx}
\usepackage{algpseudocode}

\algrenewcommand\alglinenumber[1]{{\sf\scriptsize\textcolor{LineNumGray}{\texttt{#1}}}}

\usepackage{acronym}
\acrodef{IGID}[image generation/description]{image generation/description}

\usepackage{csquotes}
\usepackage{siunitx}
\usepackage{graphicx}
\usepackage{array,booktabs,arydshln,xcolor}

\usepackage{booktabs}
\usepackage{lipsum}
\usepackage{bbm}
\usepackage{arydshln}
\usepackage{amsthm}
\usepackage{mathtools}
\usepackage{comment}
\usepackage{tikz}
\usetikzlibrary{arrows.meta,positioning}

\usepackage{url}
\usepackage{amsmath,amsfonts}
\usepackage{graphicx}
\usepackage{textcomp}
\usepackage{xcolor}
\usepackage{comment}
\usepackage{centernot} 
\usepackage{placeins}

\begin{document}

\title[Investigating Associational Biases in Inter-Model Communication of Large Generative Models]{Investigating Associational Biases in Inter-Model Communication of Large Generative Models}

\author{Fethiye Irmak Dogan}
\authornote{Corresponding Author.}
\orcid{0000-0002-1733-7019}
\email{fid21@cam.ac.uk}
\affiliation{%
  \institution{University of Cambridge}
  \city{Cambridge}
  \country{UK}
}

\author{Yuval Weiss}
\email{yw580@cam.ac.uk}
\affiliation{%
  \institution{University of Cambridge}
  \city{Cambridge}
  \country{UK}
}

\author{Kajal Patel}
\authornote{Contributed to this work while undertaking a remote visiting research studentship at the AFAR Lab, University of Cambridge, UK.}
\orcid{0009-0002-9783-4214}
\email{kpate457@illinois.edu}
\affiliation{%
  \institution{University of Illinois Urbana-Champaign}
  \city{Urbana-Champaign}
  \state{Illinois}
  \country{USA}
}

\author{Jiaee Cheong}
\orcid{0000-0001-5964-2284}
\email{jc2208@cam.ac.uk}
\affiliation{%
  \institution{University of Cambridge}
  \city{Cambridge}
  \country{UK}
}
\affiliation{%
  \institution{Harvard University}
  \city{Cambridge}
 \state{Massachusetts}
  \country{USA}
}

\author{Hatice Gunes}
\orcid{0000-0003-2407-3012}
\email{hg410@cam.ac.uk}
\affiliation{%
  \institution{University of Cambridge}
  \city{Cambridge}
  \country{UK}
}

\renewcommand{\shortauthors}{F. I. Dogan, Y. Weiss, K. Patel, J. Cheong, and H. Gunes}


\begin{abstract}

Social bias in generative AI can manifest not only as performance disparities but also as ``associational bias'', whereby models learn and reproduce stereotypical associations between concepts and demographic groups, even in the absence of explicit demographic information (e.g., associating doctors with men). 
These associations can persist, propagate, and potentially amplify across repeated exchanges in inter-model communication pipelines, where one generative model's output becomes another's input. 
This concern is especially salient for human-centred perception tasks, such as human activity recognition and affect prediction, where inferences about behaviour and internal states can lead to errors or stereotypical associations that propagate into unequal treatment in sensitive deployments (e.g., wellbeing assessment or safety monitoring). 
In this work, we focus on concepts related to human activity and affective expression, and study how such associations evolve within an inter-model communication pipeline that alternates between image generation and image description. Using the RAF-DB and PHASE datasets, we quantify demographic distribution drift induced by model-to-model information exchange and assess whether these drifts are systematic using an explainability pipeline.
Our results reveal demographic drifts toward younger representations for both actions and emotions, as well as toward more female-presenting representations, primarily for emotions. We further find evidence that some predictions are supported by spurious visual regions (e.g., background or hair) rather than concept-relevant cues (e.g., body or face). We also examine whether these demographic drifts translate into measurable differences in downstream behaviour, i.e., while predicting activity and emotion labels. Finally, we outline mitigation strategies spanning data-centric, training-time, and deployment-time (post-training) interventions, and emphasise the need for careful safeguards when deploying interconnected models in human-centred AI systems. The code is publicly available at \href{https://github.com/Cambridge-AFAR/Associational\_Biases}{https://github.com/Cambridge-AFAR/Associational\_Biases}.
\end{abstract}

\maketitle

\section{Introduction}

Associational bias in AI systems can be understood as the tendency to reproduce or strengthen social stereotypes by associating particular traits, roles, or behaviours with specific demographic groups (e.g., linking doctors with men and nurses with women)~\shortcite{mehrabi2021survey,bolukbasi_man_2016,zhao2018gender}. Such biases are not only representational: when AI outputs are used for decisions or judgements about people, stereotype-driven associations can translate into unequal system behaviour and harmful downstream outcomes~\shortcite{birhane_large_2021}.
This concern is especially important for human-centred perception tasks that span a range of difficulty, from comparatively concrete tasks such as human activity recognition, where actions can often be inferred from bodily motion and scene context, to more nuanced tasks such as affect prediction, where visual cues are more subtle and consistent annotator agreement is more challenging.
When AI systems are used for such tasks, particularly in sensitive settings (e.g., wellbeing assessment, safety monitoring, and human-centred decision support \shortcite{cheong2024small,cheong2024fairrefuse,hassan2025real}), 
stereotypical associations can be especially harmful, as they risk producing unfair or distorted interpretations of people’s behaviour.

With the rapid rise of generative AI, this concern becomes more pressing, as these models learn from large-scale, web-derived corpora that may be biased~\shortcite{bender_dangers_2021}. Indeed, biases in these models are shown to surface in model embeddings and hence in model outputs~\shortcite{basta-etal-2019-evaluating,bolukbasi_man_2016,mikolov_efficient_2013,manzini_black_2019,kotek2023gender,lee_detecting_2024}, and particularly reported for large image-text datasets~\shortcite{torralba_80_2008,birhane_large_2021} used for training vision-language and text-to-image systems. 
%
A growing body of work shows that even ordinary prompts can elicit stereotyped generations or biased predictions across demographic groups~\shortcite{seshadri2024bias, elsharif2025cultural, contreras2025automated, wan2024survey, hirota2022quantifying, chen2024would}. 
This concern is particularly acute in real-world deployments that rely on interconnected generative AI components, where the output of one model becomes the input to another, allowing biases present in individual components to propagate, interact, or evolve through inter-model information exchange.


\begin{figure}
    \centering
    \includegraphics[width=1\linewidth]{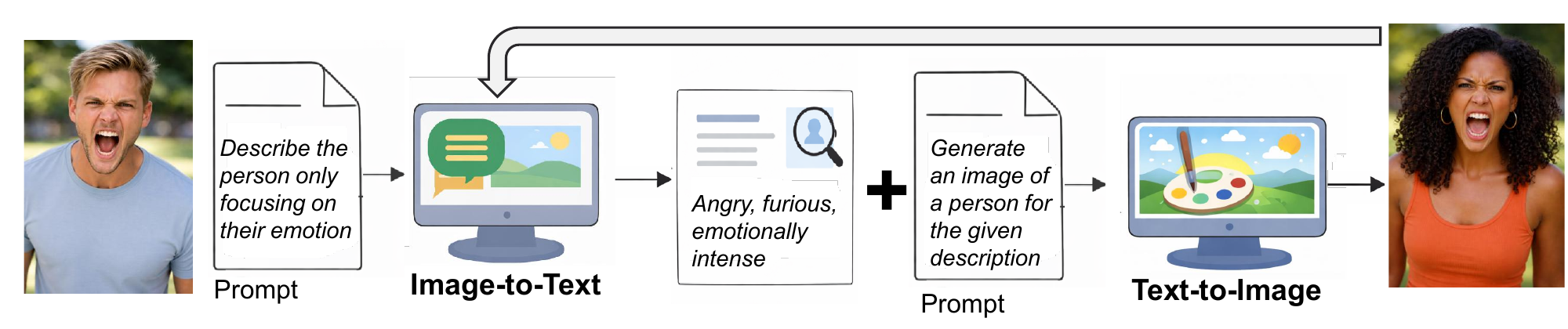}
    \caption{An illustrative example demonstrating how associational biases can arise through the inter-model communication of large generative models.}
    \label{fig:illustrate}
\end{figure}

%
Despite their importance, current bias evaluations rarely examine how demographic skew evolves through repeated model-to-model information exchange in human-centred AI settings, particularly for tasks involving human activity and affect recognition. Moreover, much of the fairness literature in these contexts has focused on predictive notions of fairness, formalising disparities via measurable criteria such as equalised odds or counterfactual fairness \shortcite{hitchhiker,cheong2023counterfactual,cheong2023causal,u-fair_ml4h_2024,cheong2024small,alam2020ai, mennella2024promoting}.
While valuable, these metrics do not directly characterise \emph{associational bias}: the tendency for particular activities or emotion labels to become disproportionately linked with specific demographic groups, even when demographic attributes are not explicitly specified \-— for example, associations between Sports and male-presenting individuals \shortcite{harrison-etal-2023-run,girrbach_large_2025}, or between Happiness and female-presenting individuals \cite{hosseini_faces_2025}.
%
%
Motivated by these gaps, we pose three research questions:
\textbf{RQ1:} How do demographic distributions drift under iterative information exchange between generative models, and do such drifts reflect stereotypical associations?
\textbf{RQ2:} When demographic drifts occur, are they attributable to sampling noise, or do they instead reflect systematic mechanisms (e.g., consistent reliance on spurious visual regions such as background or hair rather than concept-relevant cues related to human activity or affective expression)?
\textbf{RQ3:} Do these demographic drifts result in measurable differences in downstream task performance (e.g., emotion or activity prediction success) across demographic groups?

To investigate these research questions in the contexts of human activity and emotion recognition, we make four contributions:
\textbf{C1 (Demographic Distribution Drift):} We introduce an inter-model communication pipeline that couples an image-to-text describer with a text-to-image generator in two variants (text-seeded and image-seeded), enabling systematic measurement of how perceived demographic distributions drift through iterative model-to-model knowledge exchange (addresses RQ1); see Figure~\ref{fig:illustrate} for an illustrative example.
\textbf{C2 (Regional Associations):} We propose an explainability pipeline that combines token-conditioned saliency with region-based aggregation to assess whether observed demographic drifts arise from systematic mechanisms rather than incidental variation (addresses RQ2).
\textbf{C3 (Prediction Success):} We evaluate emotion and activity prediction success rates across demographic groups to quantify group-level disparities following inter-model information exchange (addresses RQ3).
\textbf{C4 (Towards Mitigation):} Finally, building on our empirical findings and prior mitigation literature, we outline mitigation directions spanning data-centric, training-time, and deployment-time (post-training) interventions for inter-model communication systems.
Across RAF-DB~\shortcite{shan2017reliable} and PHASE~\shortcite{garcia_uncurated_2023}, we observe demographic skews in loop-generated outputs, often trending toward younger representations (for both activities and emotions) and more female-presenting representations (for emotions), alongside evidence of spurious grounding. Together, these findings highlight the risks that arise when generative models are integrated as interconnected components in human-centred AI systems.
\section{Related Work}






\subsection{Biases in Human-centred AI Systems} 

\label{Bias_AI}

Biases in human-centred AI systems are a longstanding concern because even subtle biases can amplify into unfair outcomes, often disproportionately impacting marginalised groups and reinforcing societal inequities~\cite{s21217278}. Such biases can arise from imbalances in training data and surface in model outputs, affecting deployments in critical applications, including healthcare, safety monitoring, and wellbeing assessments~\cite{cheong2024small,cheong2024fairrefuse,hassan2025real,11217833}. Empirical evidence of these issues, alongside efforts to detect and mitigate them, has been documented across multiple settings, including human activity recognition~\shortcite{lai2021capturing}, facial expression recognition~\shortcite{cheong2023counterfactual,cheong2023causal}, depression identification~\shortcite{cheong2023towards,cheong2024fairrefuse,u-fair_ml4h_2024}, and pain detection~\shortcite{green2025gender}.

In response to concerns about bias in human-centred AI systems, a substantial body of work has assessed these biases primarily through predictive fairness evaluations, using formal measures such as group fairness and individual fairness~\shortcite{hort2024bias,mehrabi2021survey,hitchhiker}. Group fairness metrics typically enforce statistical parity or error-rate constraints across demographic groups, whereas individual fairness aims to ensure that similar individuals receive similar predictions~\shortcite{hort2024bias,mehrabi2021survey}.

While prior work has yielded important insights into unequal predictive performance, it largely treats bias as a property of decision outcomes~\shortcite{hort2024bias,mehrabi2021survey} rather than of representational or associational structure. In practice, human-centred AI systems can also exhibit stereotypical associations, where particular activities or emotions become disproportionately linked with specific demographics—for example, associating cooking activities or happiness with female-presenting individuals~\shortcite{zhao-etal-2017-men,hosseini_faces_2025} or linking sports activities with males~\shortcite{harrison-etal-2023-run}. Despite their potential impact, such stereotype-driven associations remain comparatively under-examined in human-centred AI, and may surface even when demographic attributes are not explicitly specified or when outcome-focused fairness metrics show limited disparity. To address this gap, our work targets stereotype-driven associational biases directly by analysing how they emerge and evolve within human-centred perception pipelines.

\subsection{Biases in Large Generative Models}

\begin{table}[t!]
\caption{Kotek et al.'s paradigm for probing gender bias in LLMs~\shortcite{kotek2023gender}.}
    \centering
    \begin{tabular}{p{\dimexpr0.12\linewidth-4\tabcolsep\relax}p{0.88\linewidth}}
    \toprule
    \multicolumn{2}{c}{A 2x2 prompt schema for probing gender bias in LLMs} \\
    \midrule
    (1) & In the sentence, `the \textbf{doctor} phoned the \textbf{nurse} because \textit{she} was late', who was late? \\[1ex]
    (2) & In the sentence, `the \textbf{nurse} phoned the \textbf{doctor} because \textit{she} was late', who was late? \\[1ex]
    (3) & In the sentence, `the \textbf{doctor} phoned the \textbf{nurse} because \textit{he} was late', who was late? \\[1ex]
    (4) & In the sentence, `the \textbf{nurse} phoned the \textbf{doctor} because \textit{he} was late', who was late? \\
    \bottomrule
    \end{tabular}
    \label{tab:wino_bias}
\end{table}

\begin{table}[t!]
\caption{Kamruzzaman et al.'s prompt schema for probing global vs local bias in LLMs \shortcite{kamruzzaman_global_2024}.}
    \centering
    \begin{tabular}{p{\dimexpr0.12\linewidth-4\tabcolsep\relax}p{0.88\linewidth}}
    \toprule
    \multicolumn{2}{c}{Schema: fill in the BLANK with the appropriate option} \\
    \midrule
    \multicolumn{2}{c}{The Nike (a global brand) shoes I had were BLANK.} \\[1ex]
    (1) & Unfashionable \\
    (2) & Fashionable \\
    (3) & Alright \\
    \bottomrule
    \end{tabular}
    \label{tab:global_local_bias}
\end{table}

Large generative models, e.g., large language models (LLMs) and vision-language models (VLMs), have been found to reproduce and even amplify biases in their outputs~\shortcite{kotek2023gender,gallegos-etal-2024-bias,naseh_backdooring_2024,kamruzzaman_global_2024,guo2024bias,seshadri2024bias,chen2024would,bianchi2023easily,hao2024harm,mandal2024generated,elsharif2025cultural,wan2024survey,hirota2022quantifying,naik2023social}. Trained on large-scale datasets, these models can inherit stereotypical patterns that surface even under neutral prompting conditions. For instance, large multimodal models systematically reproduce occupational gender stereotypes in synthetic images, often amplifying real-world imbalances across professions and defaulting to male representations~\shortcite{contreras2025automated}. Vision-language models have also been shown to produce demographically skewed generations, with White individuals making up at least 70\% of generated images in one reported analysis~\shortcite{naik2023social}. More broadly, stereotypes can be elicited by prompts referencing traits~\shortcite{seshadri2024bias}, occupations~\shortcite{elsharif2025cultural,contreras2025automated}, objects~\shortcite{wan2024survey}, or descriptors~\shortcite{hirota2022quantifying,chen2024would}.

To interpret where such behaviours originate, prior work often distinguishes between \textit{intrinsic} and \textit{extrinsic} biases~\shortcite{guo2024bias}.
Intrinsic bias can be understood as bias that stems from the underlying training data, which subsequently reinforces certain stereotypes~\shortcite{bolukbasi_man_2016,zhao2018gender}. On the other hand, extrinsic bias typically refers to bias that arises when models are fine-tuned for a specific downstream task, e.g. depression detection \shortcite{junias2025assessing,spitale2024underneath}. Within the context of this work, we investigate the \textit{associational}-biases similar to the works done in \shortcite{kotek2023gender,zhao2018gender,kamruzzaman_global_2024}. Within Kotek et al.'s paradigm~\shortcite{kotek2023gender}, each data point includes two gender-nonspecific occupations, one stereotypically associated with males and the other with females. This is then paired with a masculine or feminine pronoun, as seen in Table~\ref{tab:wino_bias}. Meanwhile, Kamruzzaman et al.~\shortcite{kamruzzaman_global_2024} analysed the presence of global vs local brand bias using a schema
illustrated in Table~\ref{tab:global_local_bias}.

Despite these advances, the growing use of generative AI in human-centred applications requires further research on how social stereotypes and associational biases emerge, propagate, and evolve across connected AI components, especially in settings where models exchange information with one another, and bias may propagate across different stages of the pipeline. To our knowledge, existing studies have not systematically investigated the associational bias that can arise from such inter-model dynamics — an important gap given the increasing use of interconnected generative pipelines in human-centred applications. Our paper addresses this gap by introducing associational bias propagation in interconnected systems and examining how biases are amplified when one model's output becomes another's input, particularly in image-to-text and text-to-image pipelines.

\subsection{Identifying Associational Biases for AI Systems}

Building on the observation in Section~\ref{Bias_AI} that stereotype-driven associational biases may not be fully captured by outcome-focused disparity evaluations~\cite{hort2024bias,mehrabi2021survey}, \emph{distributional} and \emph{explanatory} analyses can provide complementary tools for examining how such associations manifest in generated or inferred model outputs. 
In particular, distributional analyses of model outputs~\shortcite{moon2019target}, e.g., systematic drifts in the demographic composition of outputs relative to a reference, can serve as a quantitative indicator of potential associational bias. However, drift measurements alone can remain descriptive and do not explain the mechanisms behind the drift, motivating the use of explainability methods to identify which features or regions drive model behaviour. Methods such as Grad-CAM~\shortcite{selvaraju_gradcam_2017} and LIME~\shortcite{ribeiro2016should} can provide such evidence by attributing model decisions to influential input regions or features; in particular, Grad-CAM~\shortcite{selvaraju_gradcam_2017} computes the gradients of the target class with respect to the final convolutional layer
and uses these gradients to weight the feature map, generating a heatmap that highlights the most influential
regions for the model's decision~\shortcite{selvaraju_gradcam_2017}. These techniques can allow researchers to understand whether models are focusing
on relevant features (such as facial expressions in emotion recognition) or relying on irrelevant or biased cues
(e.g., background details or demographic stereotypes)~\shortcite{gebele2024interpreting, alam2024image, 10.3389/frobt.2022.937772, 9900586}.

Demographic distribution drift and explainability are particularly relevant in settings where models are composed into pipelines and one model's output becomes another model's input, as in inter-model communication. In such pipelines, drift can localise \emph{where} demographic shifts emerge across stages, while explainability can provide evidence for \emph{why} a shift may be occurring at a specific inference point by revealing the cues driving predictions~\cite{8631448}. In our setting, we use demographic distribution drift to quantify changes in demographic composition during model information exchange, and apply Grad-CAM explainability~\shortcite{selvaraju_gradcam_2017} to examine whether these shifts align with attention to semantically meaningful regions versus spurious, demographically correlated cues. \textbf{To our knowledge, this is the first work to jointly quantify demographic drift and provide attribution-based evidence for its underlying cues in an inter-model pipeline studying demographic associations with affective and emotional states.}

\section{Methodology}

\begin{figure}[t!]
    \centering
    \includegraphics[width=0.98\linewidth]{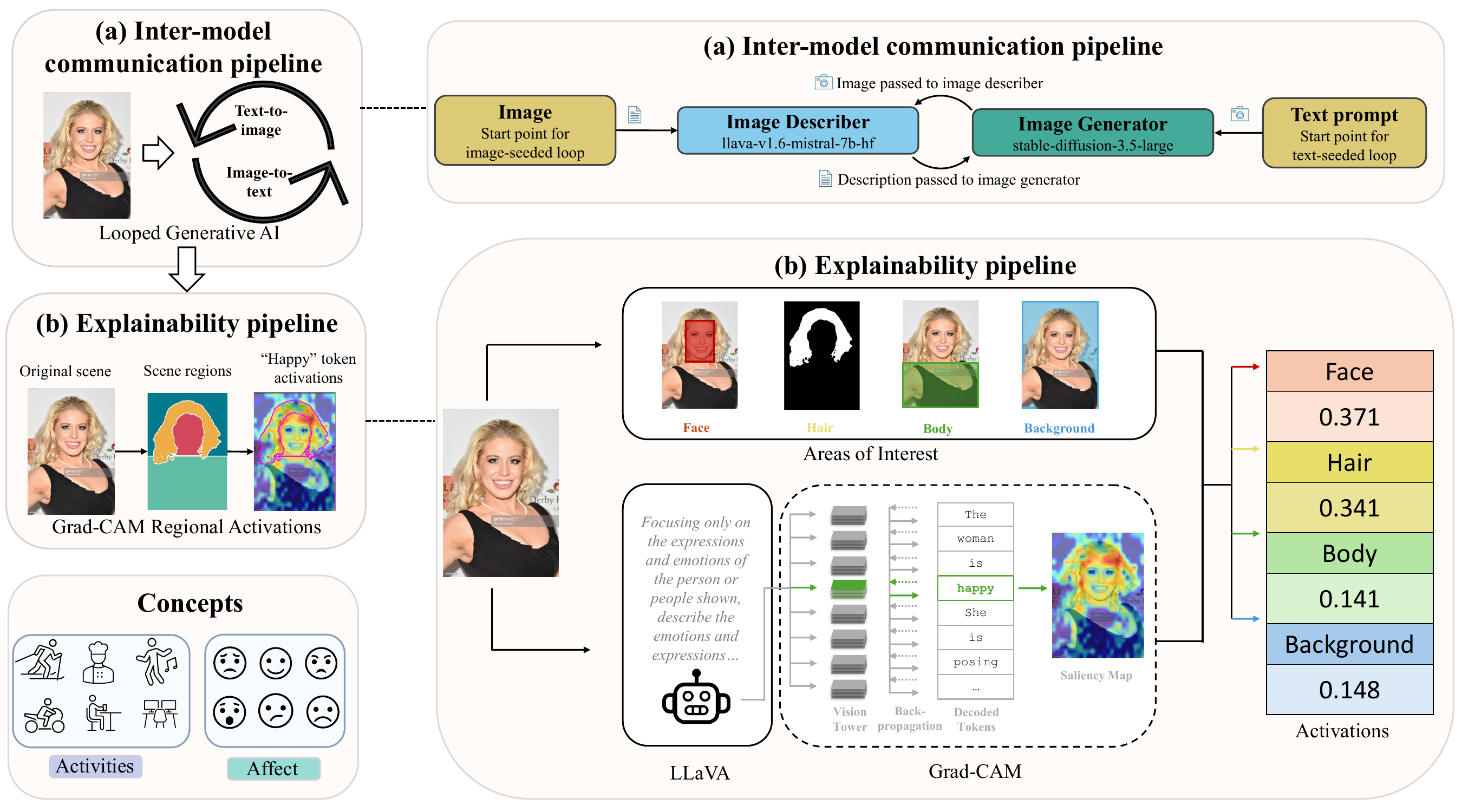}
    \caption{Overview of inter-model communication and explainability pipelines. (a) The inter-model communication pipeline runs through image generation/image description loops. (b) The explanability pipeline uses image descriptions obtained from the LLaVa model and leverages Grad-CAM to output regional activation.}
    \label{fig:pipeline_V1}
\end{figure}

In this section, we introduce two complementary pipelines designed to study how demographic distributions drift under model-to-model information exchange and to examine if this is caused by sampling noise or is systematic. First, to test whether starting conditions (e.g., different activities and affect) cause any demographic distributional drifts in generated samples, we propose an \textbf{inter-model communication pipeline} that couples a text-to-image generator with an image-to-text model in an iterative feedback loop, implemented in two variants (\emph{text-seeded} and \emph{image-seeded}). 
Second, to examine the reasons for the demographic drifts observed through inter-model communication, we propose an \textbf{explainability pipeline} that enables region-level analysis of the visual evidence on which the image-to-text model relies when generating its outputs. Specifically, we analyse regional activations during inter-model communication by computing token-conditioned saliency maps and aggregating attention over semantically meaningful regions (e.g., face, hair, body, and background). See Figure~\ref{fig:pipeline_V1} for an overview.

\subsection{Inter-model communication pipeline}

We propose an inter-model communication pipeline that enables iterative interaction between two generative models: an image-to-text model and a text-to-image model. Each model's output is used as the other's input, forming a continuous feedback loop (Figure~\ref{fig:pipeline_V1} (a)). Concretely, we consider two variants of this procedure—a text-seeded loop and an image-seeded loop—as outlined in Algorithm~\ref{algo_loop}.
\subsubsection{Mechanism 1: Text-seeded loop} \label{subsec:text_seeded}

The text-seeded loop comprises two stages that alternate between a text-to-image generator and an image-to-text describer. We begin by specifying a high-level target concept $c$ (i.e., an emotion or an activity) using a template prompt such as ``a person $y$'', where $y$ denotes an admissible label (e.g., ``doing sports'' or ``feeling happy''). This prompt $P_0$ initiates the loop and yields an image $\mathit{im}_{0}$ from the text-to-image model. Next, the generated image $\mathit{im}_{t}$ is provided to the image-to-text model using the prompt $P_{\mathrm{T}}$:
\begin{displayquote}
    \textit{Focussing on $c$ (e.g., activity or affect) of a person, and ignoring other aspects such as the fact that the image is a closeup, describe $c$ of the person in the image. Keep the description to at most 50-60 words.}
\end{displayquote}
\noindent
The resulting description $d_{t}$ is then used as the prompt for the next image generation step, producing $\mathit{im}_{t+1}$. This alternating procedure continues for $n$ iterations or until convergence. We determine convergence by computing the cosine similarity between sentence-level embeddings of consecutive descriptions:
\begin{equation}
     \cos\!\big(d_{t}, d_{t-1}) > \epsilon.
     \label{eq:convergence}
\end{equation}
When the criterion in Equation~\ref{eq:convergence} is satisfied, the looping process terminates.


\subsubsection{Mechanism 2: Image-seeded loop} \label{subsec:image_seeded}

Similar to the text-seeded loop, the image-seeded loop alternates between an image-to-text describer and a text-to-image generator. However, instead of starting from an initial text prompt, we initialise the loop with a seed image $\mathit{im}_0$ that depicts a target concept $c$ (e.g., activity or affect). 

The seed image is first provided to the image-to-text model using the prompt explained in Section~\ref{subsec:text_seeded}, yielding an initial description $d_0$. This description is then used to prompt the text-to-image model to generate the next image $\mathit{im}_1$, after which the alternating procedure continues for $n$ iterations or until convergence. To determine convergence, we compute the cosine similarity between the embeddings of consecutive generated images:
\begin{equation}
     \cos\!\big(\mathit{im}_t, \mathit{im}_{t-1}\big) > \gamma.
     \label{eq:convergence_image}
\end{equation}
When the criterion in Equation~\ref{eq:convergence_image} is satisfied, the looping process terminates.

%


\begin{algorithm}[ht!]
\footnotesize
\caption{Iterative inter-model communication pipeline.}
\label{algo_loop}
\begin{algorithmic}[1]

\Procedure{TextSeededLoop}{$c_i$}
    \State $P_0 \gets$  ``a person $c_i$'', e.g., ``a person feeling happy/doing sports'' \Comment{initial prompt}
    \State $\mathit{im}_0 \gets$ \Call{GenerateImage}{$P_0$} \Comment{text-to-image}
    \State images $\gets \{\mathit{im}_0\}$, descriptions $\gets \varnothing$
    \For{$t = 1,2,\dots$}
        \State $d_t \gets$ \Call{DescribeImage}{$P_{\mathrm{T}}, \mathit{im}_{t-1}$} \Comment{image-to-text}
        \State $\mathit{im}_t \gets$ \Call{GenerateImage}{$d_t$} \Comment{text-to-image}
        \State \textsc{append} $d_t$ to descriptions
        \State \textsc{append} $\mathit{im}_t$ to images
        \If{$\cos(d_t, d_{t-1}) > \epsilon$}
            \State \textbf{break}
        \EndIf
    \EndFor
    \State \Return images, descriptions
\EndProcedure

\Statex

\Procedure{ImageSeededLoop}{$\mathit{im}_0$}
    \State images $\gets \{\mathit{im}_0\}$, descriptions $\gets \varnothing$
    \State $d_0 \gets$ \Call{DescribeImage}{$P_{\mathrm{T}}, \mathit{im}_0$} \Comment{image-to-text}
    \State $\mathit{im}_1 \gets$ \Call{GenerateImage}{$d_0$} \Comment{text-to-image}
    \State \textsc{append} $d_0$ to descriptions
    \State \textsc{append} $\mathit{im}_1$ to images
    \For{$t = 2,3,\dots$}
        \State $d_t \gets$ \Call{DescribeImage}{$P_{\mathrm{T}}, \mathit{im}_{t-1}$} \Comment{image-to-text}
        \State $\mathit{im}_t \gets$ \Call{GenerateImage}{$d_t$} \Comment{text-to-image}
        \State \textsc{append} $d_t$ to descriptions
        \State \textsc{append} $\mathit{im}_t$ to images
        \If{$\cos(\mathit{im}_t, \mathit{im}_{t-1}) > \gamma$}
            \State \textbf{break}
        \EndIf
    \EndFor
    \State \Return images, descriptions
\EndProcedure

\end{algorithmic}
\end{algorithm}

\subsection{Explainability pipeline}
\label{exp_pipeline}
The inter-model communication pipeline allows us to generate controlled sets of images and descriptions in which demographic drifts can be quantified, but it does not explain \emph{why} such drifts may arise. We therefore introduce a complementary explainability pipeline that probes the visual evidence underpinning the image-to-text model's predictions for generating an admissible label~$y$ for a target concept~$c$ (e.g., for ``activity'' concept, $y$ can be `sports', `eating', etc.). Specifically, this pipeline localises which parts of an image most strongly support the model's selected label $y$ by computing token-conditioned saliency maps via adapting Grad-CAM~\shortcite{selvaraju_gradcam_2017} to a vision-language decoder setting and aggregating activations over semantically meaningful regions (face, hair, body, and background). This enables linking observed demographic drifts to systematic regional analyses that inform the model's decisions -- see Figure~\ref{fig:pipeline_V1} (b) for an overview.

\subsubsection{Token-conditioned saliency map}

For each image produced by the inter-model communication pipeline, we compute a token-conditioned saliency map showing the visual reasons for generating the admissible label $y$. First, we prompt the image-to-text model with a modified version of the prompt described in Section~\ref{subsec:text_seeded}, depending on whether $c$ is an activity or affect. For affect, we use the following prompt:
\begin{displayquote}
    \textit{Focusing only on the expressions and emotions of the person or people shown, describe the emotions and expressions of the person or people in the image from one of the following emotions: [happiness, sadness, fear, anger, neutral, unsure]. Keep the description to at most 50-60 words.}
\end{displayquote}
\noindent
On the other hand, for activities, the following prompt is utilised:
\begin{displayquote}
    \textit{Focusing only on the activities that the person or people shown are doing, describe the activities that the person or people shown are doing from one of the following categories of activities: [helping and caring, eating, household, dance and music, personal care, posing, sports, transportation, work, other, unsure]. Keep the description to at most 50-60 words.}
\end{displayquote}
\noindent

These prompts are used to generate a textual output, and we only retain the token corresponding to the admissible label $y$ identified for the target concept $c$, rather than the complete generated sequence. Specifically, let $\tau$ denote the position of the \emph{admissible label token} in the decoded output (how to select the admissible label token is detailed in Section~\ref{token_selection}). At position $\tau$, we define $y_{\tau}\in\mathbb{R}$ as the \emph{scalar} logit score produced by the language head for the admissible label token corresponding to $y$.

Next, let $A^k \in \mathbb{R}^{H \times W}$ denote the $k$-th vision-tower feature map (after reshaping the patch tokens into a spatial grid), where $H$ and $W$ are the height and width of this grid (i.e., the spatial resolution of the vision features). Given $y_{\tau}$ and $A^k$, Grad-CAM computes the gradient $\partial y_{\tau}/\partial A^k$ back to the vision encoder features. Specifically, it calculates channel weights by globally average-pooling~($\mathrm{GAP}$) gradients over spatial positions and forms a saliency map $L^{\tau}$:
\begin{equation}
\alpha_k^{\tau} \;=\; \mathrm{GAP}\!\left(\frac{\partial\, y_{\tau}}{\partial\, A^k}\right)
\;=\;
\frac{1}{H W}\sum_{p=1}^{H}\sum_{q=1}^{W}
\frac{\partial\, y_{\tau}}{\partial\, A^k_{pq}}\,,
\end{equation}

\begin{equation}
L^{\tau} \;=\; \mathrm{ReLU}\!\left(\sum_{k} \alpha_k^{\tau}\, A^k\right)\,.
\end{equation}

In our implementation, $A^k$ is taken from a late block of the vision encoder (specifically, layer 9) prior to multimodal fusion, thereby preserving spatial structure while encoding high-level semantic information. This ensures that the heatmaps capture the token's visual grounding before the language decoder transforms the visual features.

It is important to note that we generate the saliency maps only when the generated output contains a token corresponding to an admissible label for the concept (e.g., `happy', `sad', etc. for affect); if no admissible label token is produced (e.g., none of the valid activity/affect labels appear in the output), no saliency map is computed for that sample. Otherwise, for the selected admissible label token (at position $\tau$), we compute a token-conditioned Grad-CAM heatmap from the vision encoder, producing a spatial map at the vision-feature resolution (e.g., 24×24 for our backbone). This strategy ensures that the resulting heatmap reflects the visual evidence supporting the model’s decision for the selected admissible label, rather than being influenced by unrelated words in the generated output.

\subsubsection{Region-based aggregation and normalisation}

To convert token-conditioned saliency maps into region-level scores that summarise where the model grounds its prediction, each map $L^\tau$ is upsampled to the image resolution, and its activations are averaged over the four disjoint regions. Specifically, each image is associated with semantic regions that isolate person-level evidence: a face bounding box, a hair mask, a person/body bounding box, and a background region defined as the complement of these components. 
%

%
To obtain region-based activations, we normalise the activations for each region $r$ in an image, so that they sum to 1:
\begin{equation}
a_r^{(im)} = \frac{\text{AvgActivation}(L^\tau, r, im)}{\sum_{r' \in R} \text{AvgActivation}(L^\tau, r', im)}
\end{equation}
where $R = \{\text{hair, face, body, background}\}$, $r \in R$, and 
%
%
$\text{AvgActivation}(L^\tau, r, im)$ is the mean value of the Grad-CAM heatmap $L^\tau$ within region $r$ for image $im$. 
We then compute the mean normalised activation across all $N$ images:
\begin{equation}
\bar{a}_r = \frac{1}{N} \sum_{i=1}^{N} a_r^{(i)}\,.
\end{equation}
These values represent the average proportion of model attention (its saliency) allocated to each region, and by construction, $\sum_r \bar{a}_r \approx 1$.

\subsubsection{Token selection}
\label{token_selection}

Because saliency is computed \emph{per decoded token position} $\tau$, a single image-to-text generation can (rarely) contain multiple occurrences of admissible label tokens for the same concept (e.g., the admissible label ``happy'' may appear multiple times, or appear as part of a longer description). In the vast majority of samples, the generated output contains a single admissible label token, and we compute the token-conditioned heatmap directly for that position. When multiple admissible occurrences are present, however, an explicit selection rule is needed to avoid ambiguity about which occurrence should be treated as the model's decision. We therefore include a lightweight token-selection step that identifies a single decision-relevant position $\tau$, ensuring that the saliency map is computed only once per sample and is anchored to a consistent, decision-relevant token occurrence.

When multiple admissible label occurrences are present, we select the occurrence that functions as the explicit category \emph{prediction} in the generated output. In practice, the model may enumerate candidate categories before stating its prediction (e.g., `Out of the categories specified [helping and caring, eating, household, dance and music, personal care, posing, sports, transportation, work, other], the activity shown is sports.'). In such cases, we select the occurrence corresponding to the prediction (the second \texttt{sports} in this example), rather than earlier mentions that are part of an enumeration. Similarly, if the output is `The activity is sports. The sport they are playing is basketball.', we select the first occurrence of \texttt{sports}, as it expresses the categorical prediction.

For multi-word labels such as `Helping and Caring', we consider each constituent word independently, so either \texttt{helping} or \texttt{caring} qualifies as an admissible token. In these cases, we select the first admissible token occurring in the prediction portion of the response. This disambiguation step ensures that saliency reflects the primary classification decision rather than secondary descriptive mentions, which may correspond to different visual regions or related-but-distinct concepts. The full pipeline is shown in Figure~\ref{fig:pipeline_V1} (b) and Algorithm~\ref{algo_explainability}.

\begin{algorithm}[t!]
\footnotesize
\caption{Explainability pipeline: token-conditioned saliency, region-level aggregation and token-selection.}
\label{algo_explainability}
\begin{algorithmic}[1]

\State \textbf{Input:} $\mathcal{I}$: Set of images; $c$: Target concept; $P_{\mathrm{T}}$: Constrained prompt; $\mathcal{Y}$: Set of admissible labels for concept $c$
\State $\bar{\mathbf{a}} \gets \varnothing$ \Comment{store per-image region activations}
\ForAll{$\mathit{im} \in \mathcal{I}$}
    \State $s \gets$ \Call{GenerateOutput}{$P_{\mathrm{T}}, \mathit{im}$} \Comment{image-to-text model output string}
    \State $T \gets$ \Call{GenerateTokens}{$s$} 
    \State $T_y \gets \{t \in T \mid \Call{String}{t} \in \mathcal{Y}\}$ \Comment{retain tokens whose string matches an admissible label}
    \If{$T_y = \varnothing$}
        \State \textbf{continue} \Comment{no valid label token; skip saliency for this image}
    \EndIf
    \State $\tau \gets$ \Call{SelectDecisionToken}{$s, T_y$} \Comment{selected admissible answer's token position}
    \State $L^{\tau} \gets$ \Call{TokenGradCAM}{$\mathit{im}, \tau$} \Comment{$H \times W$ saliency map from vision features}
    \State $R_{\mathit{im}} \gets \mathcal{R}(\mathit{im})$ \Comment{obtain disjoint regions: face, hair, body (if available), background}
    \ForAll{$r \in R_{\mathit{im}}$}
        \State $u_r \gets$ \Call{AvgActivation}{$L^\tau, r, \mathit{im}$}
    \EndFor
    \ForAll{$r \in R_{\mathit{im}}$}
        \State $a_r^{(im)} \gets \frac{u_r}{\sum_{r' \in R_{\mathit{im}}} u_{r'}}
        $ \Comment{per-image normalisation across regions}
    \EndFor
    \State \textsc{append} $\{a_r^{(im)}\}_{r \in R_{\mathit{im}}}$ to $\bar{\mathbf{a}}$
\EndFor
\State \Return $\bar{a}_r = \frac{1}{|\mathcal{I}|}\sum_{\mathit{im}\in \mathcal{I}} a_r^{(\mathit{im})}$ 

\end{algorithmic}
\end{algorithm}

%

\section{Experiments}

\subsection{Datasets}


We evaluate our inter-model communication and explainability pipelines on two complementary datasets, PHASE~\shortcite{garcia_uncurated_2023} and RAF-DB~\shortcite{shan2017reliable}, which are complementary as they cover different domains and task types—controlled, i.e., in-the-wild, full-scene human activity/affect (PHASE) versus face-centric affect images (RAF-D), allowing us to test whether the observed drift and spurious grounding patterns persist across both context-rich scene cues and close-up facial cues.

First, we use \textit{PHASE} (Perceived Human Annotations for Social Evaluation)~\shortcite{garcia_uncurated_2023}, which augments a subset of images from Google Conceptual Captions with human-perceived demographic attributes and additional contextual labels, making it suitable for demographic distribution drifts, regional activation and bias analysis in broader, everyday scenes rather than only close-up faces. PHASE supports two types of target concepts in our experiments:  \emph{activities} (Helping and Caring, Eating, Household, Dance and Music, Personal Care, Posing, Sports, Transportation, Work) and \emph{emotions} (Anger, Fear, Happy, Neutral, Sad),  which serve as the target concept labels for seeding the loop and for subsequent evaluation.

Second, we \textit{RAF-DB} (Real-world Affective Faces Database)~\shortcite{shan2017reliable} that consists of unconstrained, real-world face images sourced online and annotated for facial expressions, including both ``basic'' and ``compound'' categories. In this work, we restrict our analysis to the \emph{seven basic} emotion labels provided in the dataset. Surprise, Fear, Disgust, Happiness, Sadness, Anger, and Neutral.

\subsubsection{Pre-processing}

For the explainability pipeline, we pre-processed the datasets to decompose each image into four distinct spatial regions: hair, face, body, and background. Faces were detected using RetinaFace, which outputs tight bounding boxes around visible facial regions \shortcite{deng_retinaface_2020}. 
Per detected face, we segmented hair using MediaPipe's hair segmentation model, which generates a binary mask for pixels classified as hair \shortcite{lugaresi_mediapipe_2019,tkachenka_real_2019}. 
For the Phase dataset images, body bounding boxes were taken directly from the annotations. We then adjusted these bounding boxes by setting the top boundary to align with the bottom of the face bounding box. Only the body bounding boxes associated with the activity or emotion category of interest were allowed. 
Since RAF-DB features close-up facial images without full body visibility, we do not analyse body regions for that dataset. 
%
The background region was defined to be all image pixels not assigned to hair, face, or body (for Phase), or hair and face (for RAF-DB).

\subsubsection{Category Selection} For inter-model communication experiments, we used the full datasets PHASE and RAF-DB. On the other hand, for the explanability-focused experiments, we used a small subset of action and emotion categories, where the category selections were motivated by prior work on gender representation in both behavioural and affective contexts. Specifically, we select two activities (\textit{Sports} and \textit{Caring}) from the PHASE dataset: sports and athletics are repeatedly characterised as male-coded in both social perception and model-generated portrayals~\shortcite{harrison-etal-2023-run,girrbach_large_2025}, whereas caring and caretaking roles have been shown to be strongly gendered toward women~\shortcite{girrbach_large_2025}. For emotions, we select \textit{Anger} and \textit{Happiness} from both datasets as anger has been identified as one of the most strongly stereotyped emotions in vision-language settings and in datasets that multimodal models often learn from~\shortcite{hosseini_faces_2025,dominguez-catena_less_2024}, while happiness has been reported to exhibit pronounced gender skew, frequently associating the label more with women~\shortcite{hosseini_faces_2025}. 

In terms of sensitive attribute selection (age, gender, and ethnicity), we use the full set of demographic labels provided in each dataset for the inter-model communication experiments, since these analyses are concerned with \emph{distributional drift} through the looping mechanism. For the region-based explainability analysis, we focus on gender because it is the most directly interpretable attribute in terms of visual regions (face, hair, body, background). In contrast, age and ethnicity often depend on finer-grained cues (e.g., facial structure, lighting) that are not well isolated by the existing region split, making region-level activations harder to interpret for those attributes.

\subsection{Experiments}

\paragraph{RQ1: Demographic Distribution Drift}

To answer RQ1, our main set of experiments consists of image-seeded runs (Section~\ref{subsec:image_seeded}) on PHASE (conditioned separately on activity and emotion labels) and RAF-DB (conditioned on emotion labels) datasets, using each image from these datasets as an initial input. The \acl{IGID} loop is evaluated by measuring drifts in perceived demographic attributes (gender, age, and ethnicity) between the seed inputs and the loop-generated images. We focus on image-level demographic labelling because the generated textual descriptions contain little to no explicit demographic information: the vast majority refer to the subject using gender-neutral terms (e.g., \emph{person}, \emph{they}), even when the corresponding image appears male- or female-presenting; 
mentions of ethnicity and age are similarly uncommon. Due to the scale of these datasets, it is not feasible to manually annotate the gender, age, and ethnicity of all generated images. We therefore use \texttt{llava-v1.6-mistral-7b-hf} to extract these attributes with a tightly constrained, option-only prompt schema (see Table~\ref{tab:raf_eval_schema} for the prompt schema); the outputs are post-processed to discard responses outside the admissible options. 

\begin{table}[t!]
\footnotesize
\caption{Prompt schema for evaluating the age, ethnicity, and gender of generated images (the options were selected based on the labels in the RAF-DB dataset).}
    \centering
    \begin{tabular}{p{\dimexpr0.12\linewidth-4\tabcolsep\relax}p{0.88\linewidth}}
    \toprule
    \multicolumn{2}{p{\dimexpr\linewidth-2\tabcolsep\relax}}{\raggedright Answer the following questions about the person shown in the image. Fill in the gaps based on the options provided. Do not describe the image or give extra details; only select the best option from the list.} \\
    \midrule
    1. & Ethnicity: [Choose from: Caucasian, African-American, Asian] \\
    2. & Gender: [Choose from: male, female, unsure] \\
    3. & Age: [Choose from: 0--3, 4--19, 20--39, 40--69, 70+] \\
    \bottomrule
    \end{tabular}
    \label{tab:raf_eval_schema}
\end{table}

\paragraph{RQ2: Regional Association Analysis.} To answer RQ2, we use the explanability pipeline and assess the importance (neural activations) of each scene region (face, body, background, hair, etc) attributed by the model in the image description phase, specifically while generating emotion and activity-conditioned admissible tokens. Our guiding hypothesis is that activations should align with the semantically relevant region, and the determination of what counts as `relevant' depends on the nature of the category under consideration.
In the case of emotions, the face provides the primary semantic signal. A correct attribution should therefore coincide with elevated activations in the facial region, since facial expression is the dominant marker of affect. On the other hand, for activities, 
the body and background often play central roles, as they carry the physical and contextual cues that define an activity. For instance, arms and hands may reveal helping behaviour, and playing fields or equipment provide evidence for sports activities. 


\paragraph{RQ3: Prediction Success Rates.} In addition to demographic distribution drifts and activation patterns, we measure, for each category, the proportion of instances in which the model’s \emph{selected admissible token} matches the dataset’s ground-truth activity or affect label. As described in Section~\ref{exp_pipeline}, we restrict predictions to a fixed set of admissible label tokens and count an instance as \emph{successful} if the selected token equals the target ground-truth label (e.g., \textit{happiness} for the Happiness category). We report success rates by gender and looping stage, and use them to assess fairness by comparing performance across demographic groups before and after looping.

\paragraph{Text-seeded experiments:} 
In addition to the image-seeded experiments, we ran a smaller text-seeded experiment. Using the prompt described in Section~\ref{subsec:text_seeded}, we run five \ac{IGID} iterations for each of the seven RAF-DB emotion categories. This experiment is included to probe iterative behaviour under text initialisation on a manageable scale; we report its outcomes in Appendix~\ref{app:text_seeded_results} and use it as a qualitative complement to the large-scale image-seeded runs.

\paragraph{Looping and convergence:}
To assess whether repeated \ac{IGID} iterations meaningfully change the generated content, we also analysed convergence by computing cosine similarity between consecutive iterations using sentence-embedding~\footnote{Model: \url{https://huggingface.co/sentence-transformers/all-MiniLM-L6-v2}.} similarity for descriptions and CLIP-based embedding~\shortcite{radford_learning_2021} similarity for images. Across concepts, consecutive iterations exhibit consistently high similarity, indicating that the loop stabilises quickly and additional iterations provide limited marginal change, as shown in Appendix~\ref{subsec:exploratory_analysis}. Based on this empirical convergence behaviour, our large-scale dataset experiments (RAF-DB and PHASE) use a single \ac{IGID} loop per seed image. 

\subsection{Metrics:}

To quantify \textit{RQ1}, we compare demographic distributions measured \emph{before} and \emph{after} the loop using three metrics. First, we apply the Stuart--Maxwell test of marginal homogeneity \cite{stuart_test_1955, maxwell_comparing_1970} to assess whether the pre- and post-loop categorical distributions differ significantly. To account for multiple hypothesis testing across attributes, we report p-values adjusted using the Benjamini-Hochberg false discovery rate procedure \cite{benjamini_controlling_1995}. We adjust p-values at the experimental level (i.e. Phase Activities, Phase Affect and RAF Affect). Next, we compute an unweighted Cohen's $\kappa$ score to measure the extent of agreement in demographic distributions before and after the looping, beyond what could be expected by chance. Finally, we compute a distributional similarity measure based on the weighted Jaccard index, which captures the extent of overlap between the two distributions. Specifically, we represent the pre- and post-loop distributions as non-negative vectors $\mathbf{x}$ and $\mathbf{y}$ over a common set of $n$ attribute labels and define the weighted Jaccard score as:
\begin{equation} \label{eq:jacc}
    J_\mathcal{W} = \frac{\sum_{i=1}^n \min(\mathbf{x}_i, \mathbf{y}_i)}{\sum_{i=1}^n \max(\mathbf{x}_i, \mathbf{y}_i)}.
\end{equation}

To quantify \textit{RQ2}, we evaluate the Grad-CAM-generated heatmap over the image, which we aggregate into mean activation values within the predefined regions (hair, face, body, and background). These average activations indicate where the model is focusing when generating an admissible token. Results reported based on the mean and standard deviations of activations for each region (face, body, hair, background).


To quantify \textit{RQ3}, we measure the prediction success rate for each image both before and after a single IG/ID loop. 
For each gender $g \in \{\text{male, female}\}$ and stage $s \in \{\text{before, after}\}$, we compute gender-conditioned success rates as $\text{Success Rate}_{g, s} = \frac{n_{\text{correct},g,s}}{N_{g, s}}$, where $n_{\text{correct},g,s}$ is the number of images that the model generated the correct category prediction for of gender $g$ at stage $s$, and $N_{g, s}$ is the total number of images containing that gender at that stage in the dataset. 

To further quantify the fairness of these outcomes, we focus on \emph{demographic disparity}, defined as the difference in success rates between female- and male-presenting image at a given stage: $DP_s = \text{Success Rate}_{\text{female}, s} - \text{Success Rate}_{\text{male}, s}$. We compare fairness across stages by reporting the demographic parity values before and after looping. 
To determine whether looping leads to statistically significant changes in prediction accuracy, we fit a logistic regression model at the image level with correctness as the binary outcome and including the stage in the IG/ID loop and annotated gender as predictors:
\begin{equation} \label{eq:pred_acc_log_reg}
    \text{logit}[P(\text{correct}_{i, s})] = \alpha + \beta_0 \cdot \mathbf{1}[s = \text{after}]  + \beta_1 \cdot \mathbf{1}[g_{i, s} = \text{female}]
\end{equation}
where $\text{correct}_{i, s}$ is whether the prediction for image $i$ at stage $s$ is correct, $g_{i, s}$ is the annotated gender of image $i$ at stage $s$, $\alpha$ is the baseline log-odds of correct prediction for male images before looping, $\beta_0$ measures the effect of the IG/ID loop on prediction accuracy, and $\beta_1$ captures the average difference in prediction accuracy between female- and male-presenting images aggregated across stages. We additionally report odds ratios for each predictor, showing (1) before looping vs after looping and (2) male vs female, by calculating $\exp({\beta})$. An odds ratio greater than 1 indicates increased odds of prediction success, while an odds ratio less than 1 indicates decreased odds.

\subsection{Implementation Details}
\label{sec:implementation}

All experiments instantiate the inter-model communication loop with a text-to-image generator and an image-to-text describer. For text-to-image, we use Stable Diffusion 3.5 Large \shortcite{esser_scaling_2024}. For image-to-text, we use LLaVa-Next 
with \texttt{Mistral-7B-Instruct} as the underlying language backbone. Convergence in the text-seeded loop is computed from sentence-level embeddings obtained with \texttt{all-MiniLM-L6}, while convergence in the image-seeded loop is computed from image embeddings produced by \texttt{clip-vit-large-patch14} (ViT-L/14). In the explainability pipeline, \texttt{RetinaFace (version 0.0.17)} was used for facial detection and \texttt{MediaPipe (version 0.10.21)} was utilised for hair segmentation. We implement both looping mechanisms and the explainability pipeline in \texttt{PyTorch} using the \texttt{HuggingFace} transformers and diffusers. Inter-model communication experiments were run on the University of Cambridge Research Computing Service HPC platform (Dell PowerEdge XE8545 nodes with A100-SXM-80GB GPUs running Rocky Linux 8), and explinability experiments were run on Google Colab (hosted Jupyter notebooks with GPU acceleration, specifically NVIDIA L4 GPUs running on Ubuntu 22.04).


\section{Findings} \label{sec:results}


\subsection{Findings on Demographic Distribution Drifts (RQ1)}

\label{findings-RQ1}
This section presents our findings on the PHASE and RAF-DB datasets across different activities and emotions, analysing drifts in demographic distributions through image-seeded inter-model communication.



\begin{figure*}[t!]
\centering
\includegraphics[width=0.95\linewidth]{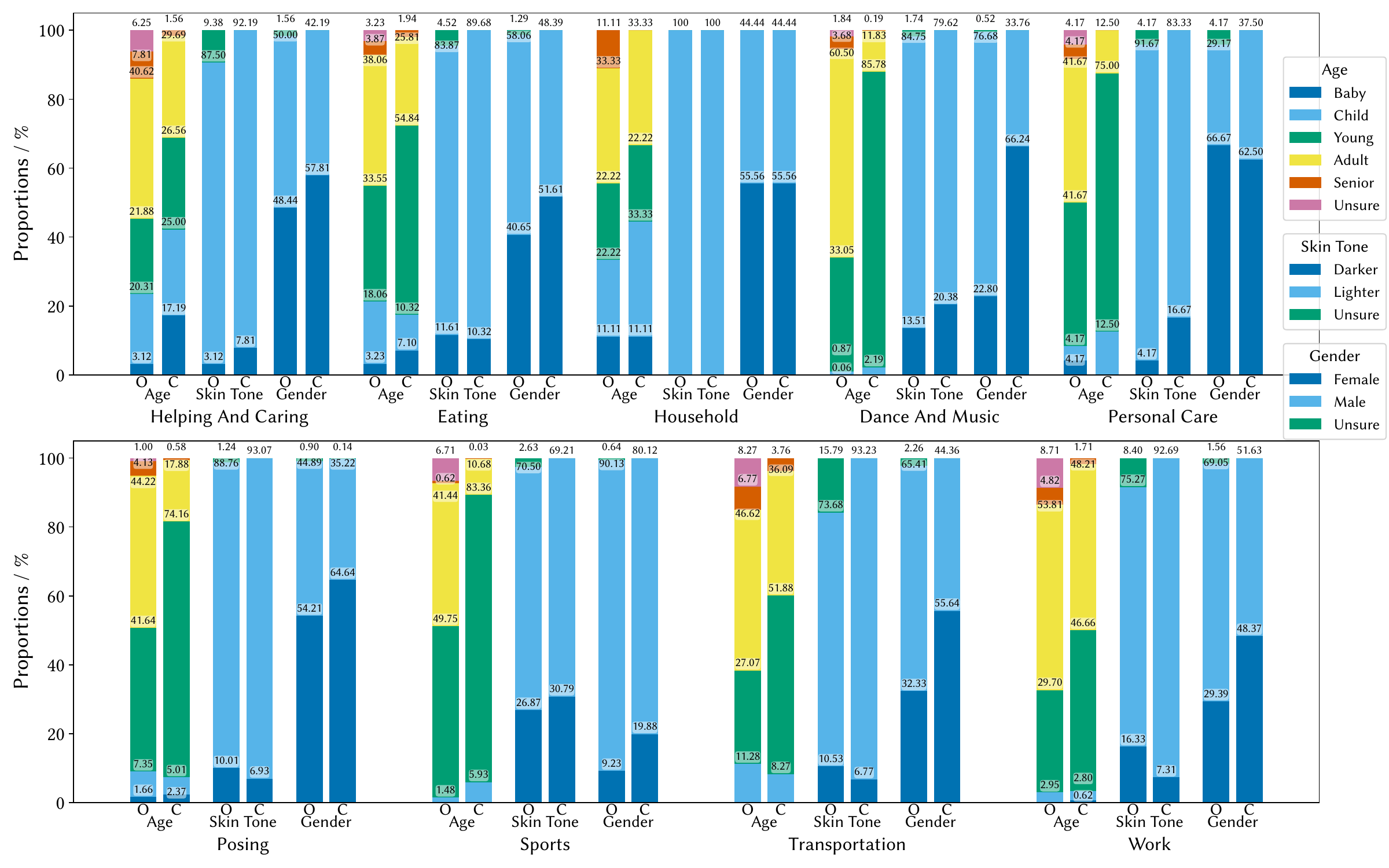}
\vspace{-1.5em}
\caption{Original (O) and collected (C) distributions for PHASE activities before and after the loop.}
\label{fig:phase_acts}
\vspace{0.7em}

\footnotesize
\captionsetup{type=table}
\caption{PHASE Activities: Stuart-Maxwell marginal homogeneity scores (bold values indicate significance under Benjamini-Hochberg with $\alpha = 0.01$; raw p-values and adjusted q-values displayed in Table~\ref{tab:phase-act-pvals}, Appendix~\ref{significance_appendix}), Cohen's $\kappa$ and Jaccard scores. }
\vspace{-1em}
\label{tab:phase-act-sim}
\begin{tabular}{
  l
  S[table-format=4.2]
  S[table-format=1.4] 
  S[table-format=1.4] 
  S[table-format=4.2]
  S[table-format=1.4] 
  S[table-format=1.4]
  S[table-format=4.2]
  S[table-format=1.4] 
  S[table-format=1.4] 
}
\toprule
\textbf{Activity} 
& \multicolumn{3}{c}{\textbf{Age}} 
& \multicolumn{3}{c}{\textbf{Skin Tone}} 
& \multicolumn{3}{c}{\textbf{Gender}} \\
\cmidrule(lr){2-4} \cmidrule(lr){5-7} \cmidrule(lr){8-10}
& {$\chi^2_\mathrm{SM}(\mathrm{df} = 4)$} & {Cohen's $\kappa$} & {Jaccard} 
& {$\chi^2_\mathrm{SM}(\mathrm{df} = 1)$} & {Cohen's $\kappa$}& {Jaccard} 
& {$\chi^2_\mathrm{SM}(\mathrm{df} = 1)$} & {Cohen's $\kappa$}& {Jaccard} \\ 
\midrule
Helping \& Caring & 7.28 & 0.0459 & 0.6533 & 1.00 & 0.3012 & 0.9062 & 1.19 & 0.3348 & 0.8406 \\
Eating & \bfseries 23.12 & 0.2808 & 0.6138 &  0.36 & 0.1483 & 0.9299 & 4.57 & 0.2722 & 0.8118 \\
Household & 0.09 & 0.4098 & 0.8000 & 0.00 & 1.0000 & 1.0000 & 0.00 & 0.1000 & 1.0000 \\
Dance \& Music & \bfseries 1485.63 & 0.0409 & 0.3019 & \bfseries 50.57 & 0.0463 & 0.8859 & \bfseries 1154.97 & 0.1428 & 0.3958 \\
Personal Care & 5.69 & 0.0828 & 0.4242 & 1.00 & -0.0698 & 0.8077 & 0.20 & 0.5064 & 0.8800 \\
Posing & \bfseries 1442.37 & 0.2100 & 0.5050 & \bfseries 41.29 & 0.1043 & 0.9282 & \bfseries 215.68 & 0.4325 & 0.8165 \\
Sports & \bfseries 993.84 & 0.1158 & 0.4716 & \bfseries 11.72 & 0.2708 & 0.9486 & \bfseries 206.80 & 0.2644 & 0.8122 \\
Transportation & \bfseries 15.33 & 0.1819 & 0.6452 & 1.32 & 0.0843 & 0.7754 & \bfseries 17.89 & 0.3024 & 0.6335 \\
Work & \bfseries 42.37 & 0.1687 & 0.7571 & \bfseries 35.12 & 0.1441 & 0.7575 & \bfseries 69.39 & 0.3278 & 0.6901 \\
\bottomrule
\end{tabular}
\vspace{-1em}
\end{figure*}

\begin{figure*}[t!]
\centering
\includegraphics[width=0.95\linewidth]{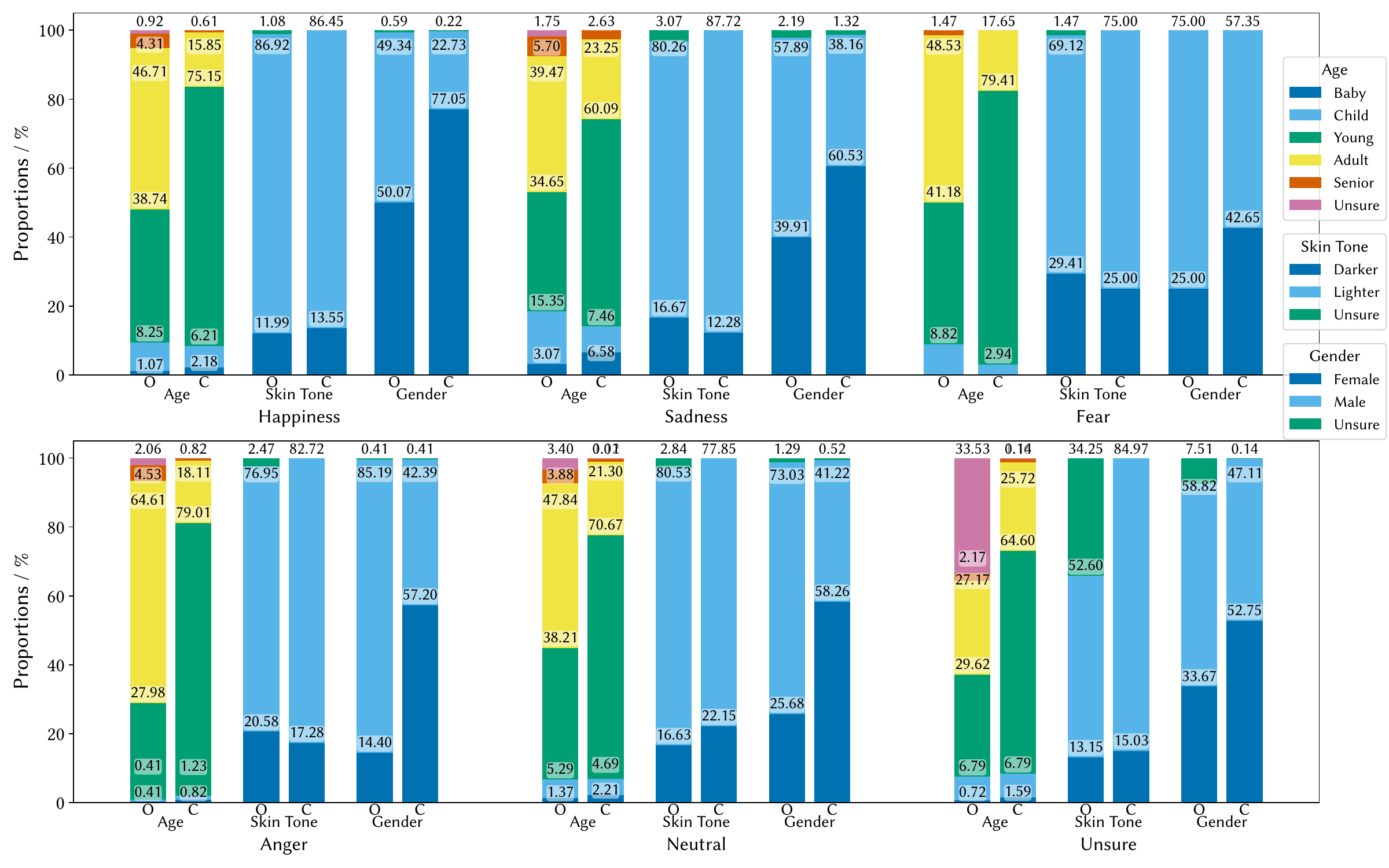}
\vspace{-1.5em}
\caption{Original (O) and collected (C) distributions for PHASE emotions before and after the loop.
}
\label{fig:phase_emos}
\vspace{0.7em}

\footnotesize
\captionsetup{type=table}
\caption{PHASE Emotions: Stuart-Maxwell marginal homogeneity scores (bold values indicate significance under Benjamini-Hochberg with $\alpha = 0.01$; raw p-values and adjusted q-values displayed in Table~\ref{tab:phase-ems-pvals}, Appendix~\ref{significance_appendix}), Cohen's $\kappa$ and Jaccard scores. }
\vspace{-1em}
\label{tab:phase-em-sim}
\begin{tabular}{
  l
  S[table-format=4.2]
  S[table-format=1.4] 
  S[table-format=1.4] 
  S[table-format=4.2]
  S[table-format=1.4] 
  S[table-format=1.4]
  S[table-format=4.2]
  S[table-format=1.4] 
  S[table-format=1.4] 
}
\toprule
\textbf{Emotion} 
& \multicolumn{3}{c}{\textbf{Age}} 
& \multicolumn{3}{c}{\textbf{Skin Tone}} 
& \multicolumn{3}{c}{\textbf{Gender}} \\
\cmidrule(lr){2-4} \cmidrule(lr){5-7} \cmidrule(lr){8-10}
& {$\chi^2_\mathrm{SM}(\mathrm{df} = 4)$} & {Cohen's $\kappa$} & {Jaccard} 
& {$\chi^2_\mathrm{SM}(\mathrm{df} = 1)$} & {Cohen's $\kappa$}& {Jaccard} 
& {$\chi^2_\mathrm{SM}(\mathrm{df} = 1)$} & {Cohen's $\kappa$}& {Jaccard} \\ 
\midrule
Anger & \bfseries 107.00 & 0.0358 & 0.3178 &  1.03 & 0.1653 & 0.9124 & \bfseries 98.33 & 0.1768 & 0.3988 \\
Fear & \bfseries 17.72 & 0.0947 & 0.4468 & 0.43 & 0.2179 & 0.9014 & 6.00 & 0.2381 & 0.7000 \\
Happy & \bfseries 1982.68 & 0.1663 & 0.4574 &  6.37 & 0.0743 & 0.9797 & \bfseries 1226.74 & 0.1813 & 0.5760 \\
Neutral & \bfseries 1923.62 & 0.1720 & 0.5134 & \bfseries 85.96 & 0.1503 & 0.9202 & \bfseries 2158.08 & 0.2084 & 0.5095 \\
Sad & \bfseries 36.47 & 0.2385 & 0.5586 & 2.88 & 0.0919 & 0.8866 & \bfseries 25.99 & 0.2667 & 0.6593 \\
\bottomrule
\end{tabular}
\vspace{-1em}
\end{figure*}

\begin{figure*}[t!]
\centering
\includegraphics[width=0.95\linewidth]
{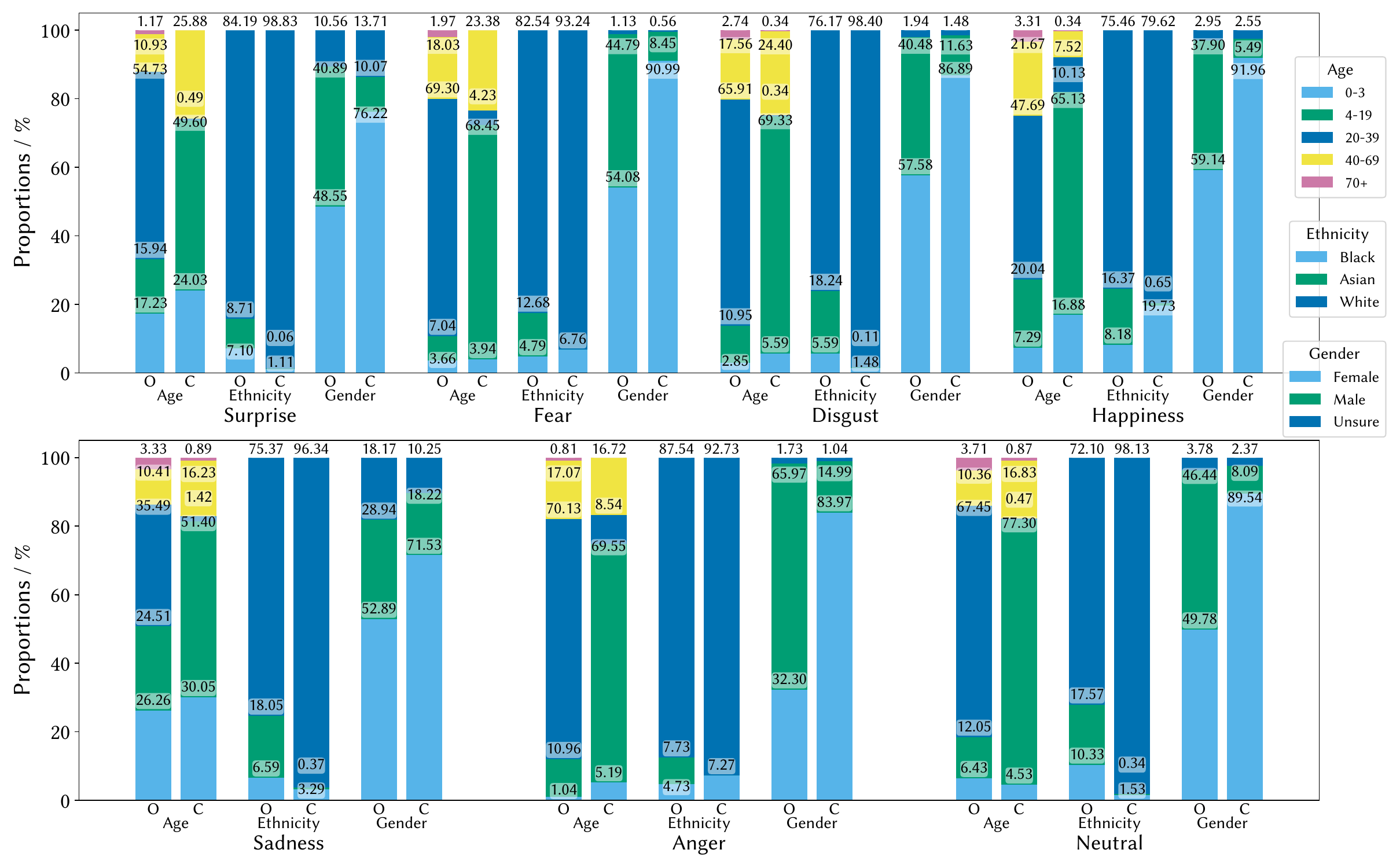}
\vspace{-1.5em}
\caption{Original (O) and collected (C) distributions for RAF-DB before and after.
}
\label{fig:whole_raf}
\vspace{0.7em}

\footnotesize
\captionsetup{type=table}
\caption{RAF-DB Emotions: Stuart-Maxwell marginal homogeneity scores (bold values indicate significance under Benjamini-Hochberg with $\alpha = 0.01$; raw p-values and adjusted q-values displayed in Table~\ref{tab:raf-pvals}, Appendix~\ref{significance_appendix}), Cohen's $\kappa$ and Jaccard scores.}
\vspace{-1em}
\label{tab:raf-sim}
\begin{tabular}{
  l
  S[table-format=4.2]
  S[table-format=1.4] 
  S[table-format=1.4] 
  S[table-format=4.2]
  S[table-format=1.4] 
  S[table-format=1.4]
  S[table-format=4.2]
  S[table-format=1.4] 
  S[table-format=1.4] 
}
\toprule
\textbf{Emotion} 
& \multicolumn{3}{c}{\textbf{Age}} 
& \multicolumn{3}{c}{\textbf{Ethnicity}} 
& \multicolumn{3}{c}{\textbf{Gender}} \\
\cmidrule(lr){2-4} \cmidrule(lr){5-7} \cmidrule(lr){8-10}
& {$\chi^2_\mathrm{SM}(\mathrm{df} = 4)$} & {Cohen's $\kappa$} & {Jaccard} 
& {$\chi^2_\mathrm{SM}(\mathrm{df} = 2)$} & {Cohen's $\kappa$}& {Jaccard} 
& {$\chi^2_\mathrm{SM}(\mathrm{df} = 2)$} & {Cohen's $\kappa$}& {Jaccard} \\
\midrule
Surprise & \bfseries 943.57 & 0.0981 & 0.2870 & \bfseries 211.06 & -0.0051 & 0.7446 & \bfseries 361.82 & 0.1590 & 0.5288 \\
Fear & \bfseries 247.54 & 0.0545 & 0.1973 & \bfseries 42.09 & 0.0330 & 0.7750 & \bfseries 104.23 & 0.0231 & 0.4609 \\
Disgust & \bfseries 640.87 & 0.0044 & 0.1908 & \bfseries 179.24 & -0.0011 & 0.6362 & \bfseries 177.14 & 0.0395 & 0.5467 \\
Happiness & \bfseries 3533.30 & 0.0223 & 0.2930 & \bfseries 1044.38 & -0.0165 & 0.7285 & \bfseries 1705.04 & 0.0857 & 0.5059 \\
Sadness & \bfseries 874.18 & 0.1480 & 0.4653 & \bfseries 466.54 & 0.0253 & 0.6534 & \bfseries 265.12 & 0.2316 & 0.6857 \\
Anger & \bfseries 583.36 & -0.0168 & 0.2289 & \bfseries 65.64 & 0.0385 & 0.8565 & \bfseries 389.60 & 0.0384 & 0.3186 \\
Neutral & \bfseries 2309.74 & 0.0652 & 0.1647 & \bfseries 776.32 & 0.0164 & 0.5871 & \bfseries 1117.69 & 0.0997 & 0.4311 \\
\bottomrule
\end{tabular}
\vspace{-1em}
\end{figure*}


\subsubsection{PHASE Activities:}
\label{subsec:phachi-square}

The demographic distribution drifts for the PHASE activity experiment before and after the \ac{IGID} loop are shown in Figure~\ref{fig:phase_acts}. Table~\ref{tab:phase-act-sim} reports (i) $\chi^2_\mathrm{SM}$ values from the Stuart-Maxwell marginal homogeneity test, adjusted using Bejamini-Hochberg false-discovery control with $\alpha$ (family-wide error rate) set to 0.01, (ii) Cohen's $\kappa$ scores indicating pairwise agreement of the pre- and post-loop distributions, if the initial and former distributions agree with one another,  (iii) weighted Jaccard similarities quantifying the magnitude of overlap between the two distributions.


In terms of age, 
the results show a trend toward younger-presenting individuals, significant for \textit{Easting}, \textit{Music and Dance}, \textit{Posing}, \textit{Sports}, \textit{Transportation} and \textit{Work}. Cohen's $\kappa$ and Jaccard similarities are the lowest between pre/post distributions for the \textit{Music and Dance} category, highlighting the largest drift of distributions based on age groups, i.e., younger representations are 33.05\% in the original dataset and increase to 85.78\% in the generated images. In terms of skin-tone, Jaccard similarities are higher, and drifts are activity-dependent, e.g., \textit{Work} had more lighter-skin presenting individuals in the post-loop distribution than in the original one; but this effect was reversed for \textit{Music and Dance}. Finally, in terms of gender, \textit{Music and Dance} showed the most significant drift with the lowest Cohen's $\kappa$ and Jaccard similarities, showing the most significant change in distribution towards females, 22.80\% in the original distribution vs 66.24\% in the generated images. For the rest of the activities, the drifts were comparatively modest.


\subsubsection{PHASE Emotions:}

The demographic distribution drifts and scores for the PHASE-emotion experiments are shown in Figure~\ref{fig:phase_emos} and Table~\ref{tab:phase-em-sim}. 
In terms of age, low Cohen's $\kappa$ and Jaccard similarities are observed (indicating limited overlap between the pre-loop and post-loop age distributions), and the results show drifts towards younger representations for all emotions. For instance, younger individuals are 20.04\% for \textit{Happiness} in the original dataset and increase to 65.13\% in the generated images. In terms of ski-tone, distribution drifts are not significant except for \textit{Neutral}, and weighted Jaccard similarities are overall the highest. For gender, these similarities are relatively lower, and the Stuart-Maxwell marginal homogeneity scores indicate that these changes are statistically significant for every emotion except \textit{fear} after the \ac{IGID} loop, presenting a drift towards female-presenting individuals, e.g., 59.14\% for \textit{Happiness} in the original distribution, which reaches to 91.96\% in the generated images. 



\subsubsection{RAF-DB Emotions}

The demographic distributions before and after the looping and their similarity scores are presented in Figure \ref{fig:whole_raf} and Table~\ref{tab:raf-sim}. 
The results highlight that distributional drifts are more pronounced with lower Cohen's $\kappa$ and Jaccard similarities than in the Phase dataset. In terms of age, across all emotions, the experimental distribution drifts towards younger-presenting people. For example, 24.98\% of images labelled as happy were of people in the 40-69 and 70+ categories, whereas in the experimental data, only 7.86\% of images were classified as such. 

Regarding ethnicity, there are significantly fewer Asian-presenting individuals in the generated images than in the original RAF-DB distribution. For example, in the original distribution, the proportions of Asian-presenting individuals for happiness and sadness were 16.37\% and 18.05\%, respectively. By contrast, these proportions were reduced to only 0.65\% and 0.37\% in the generated set.
For happiness and anger, there was an increase in the number of Black representations, with proportions rising from 8.18\% to 19.73\% and from 4.73\% to 7.27\%, respectively. 
For surprise, fear, disgust, and neutral, white-presenting individuals became more pronounced, as exemplified by the proportion of white representations in the surprised category increasing from 86.02\% to 98.18\%. Similarly, the proportion of white-presenting individuals in the disgusted category increased from 84.19\% to 98.83\%. 

Regarding gender, although the gender distribution is approximately even in the original RAF-DB dataset, the \ac{IGID} distribution heavily favours females across all the emotions. For instance, female-presenting individuals increase from 59.14\% to 91.96\% for the \textit{Happiness} emotion. 
Overall, although the magnitude differs, the direction of change mirrors Phase findings, with post-loop samples skewing more toward younger, female-presenting individuals.



\subsection{Findings on Regional Associations (RQ2)}

The previous section analyses demographic distribution drifts under inter-model communication, and in this section, we provide additional explainability insights by examining regional activations for different activities and emotions across gender groups.

\begin{table}[t!]
\footnotesize
\centering
\caption{Normalised regional activation fractions (mean ± std) for the Happiness and Anger emotions across PHASE and RAF-DB datasets.}
\begin{subtable}[t]{0.30\textwidth}
\footnotesize
\centering
\setlength{\tabcolsep}{12pt}
\renewcommand{\arraystretch}{1.25}
\caption{Phase dataset - Sports}
\label{tab:phase-sports-overall-activations}
\begin{tabular}{lc}
\toprule
\textbf{Region} & \textbf{Overall} \\
\midrule
Hair         & 0.231 $\pm$ 0.135 \\
Face         & 0.227 $\pm$ 0.150 \\
Body         & 0.273 $\pm$ 0.124 \\
Background   & 0.269 $\pm$ 0.170 \\
\bottomrule
\end{tabular}
\end{subtable}
\hfill
\begin{subtable}[t]{0.30\linewidth}
    \centering
\centering
\setlength{\tabcolsep}{12pt}
\renewcommand{\arraystretch}{1.25}
\caption{Phase dataset - Happiness}
\label{tab:phase-happiness-overall-activations}
\begin{tabular}{lc}
\toprule
\textbf{Region} & \textbf{Overall} \\
\midrule
Hair         & 0.224 $\pm$ 0.130 \\
Face         & 0.231 $\pm$ 0.174 \\
Body         & 0.270 $\pm$ 0.123 \\
Background   & 0.276 $\pm$ 0.169 \\
\bottomrule
\end{tabular}

    \label{tab:happiness_anger__activations_1}
\end{subtable}
\hfill
\begin{subtable}[t]{0.30\linewidth}
    \centering
\centering
\setlength{\tabcolsep}{12pt}
\renewcommand{\arraystretch}{1.25}
\caption{Phase dataset - Anger}
\label{tab:phase-anger-overall-activations}
\begin{tabular}{lc}
\toprule
\textbf{Region} & \textbf{Overall} \\
\midrule
Hair         & 0.234 $\pm$ 0.129 \\
Face         & 0.238 $\pm$ 0.156 \\
Body         & 0.277 $\pm$ 0.113 \\
Background   & 0.251 $\pm$ 0.149 \\
\bottomrule
\end{tabular}

    \label{tab:happiness_anger__activations_2}
\end{subtable}

\vspace{2em}
\begin{subtable}[t]{0.30\textwidth}
\footnotesize
\centering
\setlength{\tabcolsep}{12pt}
\renewcommand{\arraystretch}{1.25}
\caption{Phase dataset - Caring}
\label{tab:phase-caring-overall-activations}
\begin{tabular}{lc}
\toprule
\textbf{Region} & \textbf{Overall} \\
\midrule
Hair         & 0.231 $\pm$ 0.139 \\
Face         & 0.239 $\pm$ 0.145 \\
Body         & 0.280 $\pm$ 0.090 \\
Background   & 0.250 $\pm$ 0.123 \\
\bottomrule
\end{tabular}
\end{subtable}
\hfill
\begin{subtable}[t]{0.30\linewidth}
    \centering
\centering
\setlength{\tabcolsep}{12pt}
\renewcommand{\arraystretch}{1.25}
\caption{RAF-DB dataset - Happiness}
\label{tab:raf-db-happiness-overall-activations}
\begin{tabular}{lc}
\toprule
\textbf{Region} & \textbf{Overall} \\
\midrule
Hair         & 0.306 $\pm$ 0.152 \\
Face         & 0.320 $\pm$ 0.204 \\
Background   & 0.374 $\pm$ 0.225 \\
\bottomrule
\end{tabular}
    \label{tab:happiness_anger__activations_3}
\end{subtable}
\hfill
\begin{subtable}[t]{0.30\linewidth}
    \centering
\centering
\setlength{\tabcolsep}{12pt}
\renewcommand{\arraystretch}{1.25}
\caption{RAF-DB dataset - Anger}
\label{tab:raf-db-anger-overall-activations}
\begin{tabular}{lc}
\toprule
\textbf{Region} & \textbf{Overall} \\
\midrule
Hair         & 0.304 $\pm$ 0.153 \\
Face         & 0.342 $\pm$ 0.201 \\
Background   & 0.354 $\pm$ 0.220 \\
\bottomrule
\end{tabular}
    \label{tab:happiness_anger__activations_4}
\end{subtable}

\label{tab:happiness_anger__activations}
\vspace{-1em}
\end{table}

\subsubsection{Activities}

For the \emph{Sports} activity (Table \ref{tab:phase-sports-overall-activations}), the model shows the strongest activations on the body region (0.273 $\pm$ 0.124), followed by the background region (0.269 $\pm$ 0.170). 
This distribution suggests that the model captures contextual cues, like fields, courts, crowds, and outdoor environments, alongside body movement and posture, which is consistent with the idea that sports imagery frequently includes significant background information relating to the setting of the activity. 
In these activities, expression cues (face region activation: 0.227 $\pm$ 0.150) can be relatively less salient than the physical action and contextual setting.

For the \emph{Caring} activity (Table \ref{tab:phase-caring-overall-activations}), 
the body and background regions contributes most (0.280 $\pm$ 0.090 and 0.250 $\pm$ 0.123, respectively), followed by face (0.239 $\pm$ 0.145) and hair (0.231 $\pm$ 0.139). Here, the model places slightly more emphasis on the body and facial areas, which may reflect the interpersonal and expressive nature of caring actions, such as hugging, comforting, or assisting. The attention on the background may also capture relevant contextual information, such as an ambulance or hospital. Overall, the activations together suggest that the model interprets caring both through individual expressivity and social context.

\subsubsection{Emotions}

\begin{figure}[t!]
\centering
\begin{subfigure}{.24\textwidth}
    \centering
    \includegraphics[width=.95\linewidth]{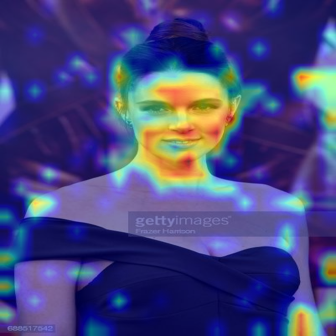}  
    \caption{Happiness - Appropriate attention}

    \label{fig:emotion_attention_examples_1}
\end{subfigure}
\begin{subfigure}{.24\textwidth}
    \centering
    \includegraphics[width=.95\linewidth]{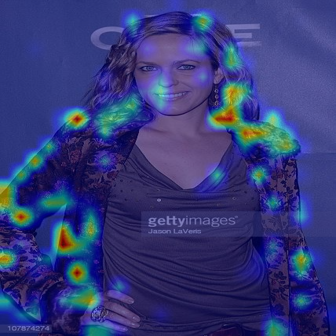}  
    \caption{Happiness - Inappropriate attention}
    \label{fig:emotion_attention_examples_2}
\end{subfigure}
\begin{subfigure}{.24\textwidth}
    \centering
    \includegraphics[width=.95\linewidth]{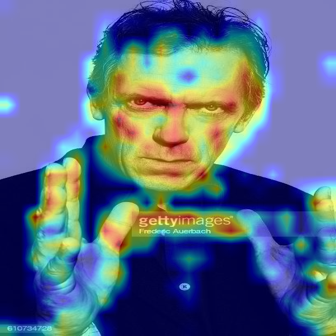}  
    \caption{Anger - Appropriate attention}
    \label{fig:emotion_attention_examples_3}
\end{subfigure}
\begin{subfigure}{.24\textwidth}
    \centering
    \includegraphics[width=.95\linewidth]{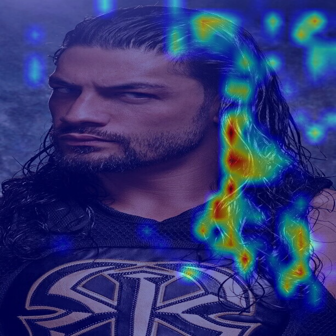}  
    \caption{Anger - Inappropriate attention}
    \label{fig:emotion_attention_examples_4}
\end{subfigure}
\vspace{-1em}
\caption{Example Grad-CAM activation heatmaps showing appropriate versus inappropriate attention patterns for happy and anger emotion recognition.}
\label{fig:emotion_attention_examples}
\vspace{-0.5cm}
\end{figure}

The results for the \emph{Happiness} emotion are shown in Table \ref{tab:happiness_anger__activations_1} and \ref{tab:happiness_anger__activations_3}. 
In PHASE, activations are relatively similarly distributed across regions, with body (0.270 $\pm$ 0.123), background (0.276 $\pm$ 0.169) contributing slightly more than face (0.231 $\pm$ 0.174), and hair (0.224 $\pm$ 0.130). In RAF-DB, the model shows slightly higher reliance on background (0.374 $\pm$ 0.225) and face (0.320 $\pm$ 0.204), with hair also contributing substantially (0.306 $\pm$ 0.152). 
%

For the \emph{Anger} emotion (Table \ref{tab:happiness_anger__activations_2} and \ref{tab:happiness_anger__activations_4}), the pattern shows more balanced attention across regions. In the PHASE dataset, body activation is highest (0.277 $\pm$ 0.113), followed by background (0.251 $\pm$ 0.149), face (0.238 $\pm$ 0.156), and hair (0.234 $\pm$ 0.129). In RAF-DB, background shows the highest prominence (0.354 $\pm$ 0.220) but is nearly equal to face (0.342 $\pm$ 0.201), followed by hair (0.304 $\pm$ 0.153).

Figure~\ref{fig:emotion_attention_examples} further illustrates representative examples of token-conditioned saliency for \emph{Happiness} and \emph{Anger}. In appropriate cases, the model’s activation is concentrated on semantically relevant evidence—most notably the face when predicting emotions (Figures~\ref{fig:emotion_attention_examples_1} and \ref{fig:emotion_attention_examples_3}). In contrast, inappropriate cases show activations dominated by less relevant or spurious cues, such as background context or hair regions (Figures~\ref{fig:emotion_attention_examples_2} and \ref{fig:emotion_attention_examples_4}).

\subsection{Findings on Prediction Success Rates (RQ3)}

\begin{table}[htbp!]
\centering
\scriptsize
\setlength{\tabcolsep}{6pt}
\renewcommand{\arraystretch}{1.2}
\caption{PHASE (\emph{Sports}, \emph{Caring}, \emph{Happiness}, \emph{Anger}) and RAF-DB (\emph{Happiness}, \emph{Anger}) prediction outcomes by gender with demographic parity (DP) and logistic regression results for prediction accuracy. Male is the reference group for ``vs Male''; therefore, Male has DP diff $=0$. Coefficients represent the effect of being in the ``Before'' IG/ID looping stage (vs. ``After'') and being Male (vs. Female) on the log-odds of prediction success.
Reported values are log-odds ($\beta$), odds ratios (OR), and $p$-values. Bolded results are statistically significant at $\alpha = 0.01$.}

\begin{tabular}{lllrrrrrr}
\toprule
& & \textbf{Gender} & \textbf{Success} & \textbf{Failure} & \textbf{Total} & \textbf{Success (\%)} & \textbf{DP diff vs Male (pp)} \\
\midrule

\multirow{6}{*}{\rotatebox{90}{Sports (PHASE)}} & \multirow{2}{*}{\rotatebox{90}{Before}}

& Male   &  2277 &  946 &  3223 &  70.65 &  0.00 \\
& & Female & 220 &   110 &  330 &  66.67 &  -3.98 \\

& \multirow{2}{*}{\rotatebox{90}{After}}

& Male   &  2408 &  457 &  2865 &  84.05 & 0.00 \\
& & Female &  538 &   173 &  711 &  75.67 &  -8.38 \\
& \multicolumn{7}{l}{\textbf{Before vs After:} $\beta_0$ (log-odds): \textbf{-0.728}, OR (odds ratios): 0.48, $p$ {$<0.001$}} \\
& \multicolumn{7}{l}{\textbf{Male vs Female:} $\beta_1$ (log-odds): \textbf{0.386}, OR (odds ratios):  1.47, $p$ {$<0.001$}} \\

\midrule

\multirow{6}{*}{\rotatebox{90}{Caring (PHASE)}} & \multirow{2}{*}{\rotatebox{90}{Before}}

& Male   &  14 &  18 &  32 &  43.75 &  0.00 \\
& & Female & 10 &  21 &  31 & 32.26 &  -11.49 \\

& \multirow{2}{*}{\rotatebox{90}{After}}

& Male   & 13 & 14 & 27  & 48.15 & 0.00 \\
& & Female & 26 &  11 & 37 & 70.27 &  22.12 \\

& \multicolumn{7}{l}{\textbf{Before vs After:} $\beta_0$ (log-odds): \textbf{-0.914}, OR (odds ratios): 0.40, $p$: {$0.005$}} \\
& \multicolumn{7}{l}{\textbf{Male vs Female:} $\beta_1$ (log-odds): -0.221, OR (odds ratios):  0.80, $p$: {$0.584$}} \\

\midrule
\multirow{6}{*}{\rotatebox{90}{Happiness (PHASE)}} & \multirow{2}{*}{\rotatebox{90}{Before}}

& Male   &  2617 &  798 &  3415 &  76.63 &   0.00 \\
& & Female & 3035 &   430 &  3465 & 87.59 &  +10.96 \\

& \multirow{2}{*}{\rotatebox{90}{After}}

& Male   & 1201 & 372 & 1573 & 76.35 &  0.00 \\
& & Female & 4331 &  1001 & 5332 & 81.23 &  +4.88 \\
& \multicolumn{7}{l}{\textbf{Before vs After:} $\beta_0$ (log-odds): \textbf{0.286}, OR (odds ratios): 1.33, $p$ {$<0.001$}} \\
& \multicolumn{7}{l}{\textbf{Male vs Female:} $\beta_1$ (log-odds): \textbf{-0.541}, OR (odds ratios):  0.58, $p$ {$<0.001$}} \\
\midrule

\multirow{6}{*}{\rotatebox{90}{Anger (PHASE)}} & \multirow{2}{*}{\rotatebox{90}{Before}}

& Male   &  82 &  125 &  207 &  39.61 & 0.00  \\
& & Female & 11 &   24 &  35 & 31.43 &  -8.18 \\

& \multirow{2}{*}{\rotatebox{90}{After}}

& Male   & 39 & 64 & 103 & 37.86 & 0.00 \\
& & Female & 50 &  89 & 139 & 35.97 &  -1.89 \\
& \multicolumn{7}{l}{\textbf{Before vs After:} $\beta_0$ (log-odds): -0.003, OR (odds ratios): 1.00, $p$: {$ 0.987$}} \\
& \multicolumn{7}{l}{\textbf{Male vs Female:} $\beta_1$ (log-odds): 0.172, OR (odds ratios):  1.19, $p$: {$0.441$}} \\
\midrule

\multirow{6}{*}{\rotatebox{90}{Happiness (RAF-DB)}} & \multirow{2}{*}{\rotatebox{90}{Before}}

& Male   &  1465 &  306 &   1771 & 82.72 &   0.00 \\
& & Female & 2362 &   428 & 2790 & 84.66 &  +1.94 \\

& \multirow{2}{*}{\rotatebox{90}{After}}

& Male   & 118 & 134 & 252 & 46.82 & 0.00 \\
& & Female & 2299 & 2010 &  4309 & 53.35 & +6.53 \\
& \multicolumn{7}{l}{\textbf{Before vs After:} $\beta$ (log-odds): \textbf{1.593}, OR (odds ratios): 4.92, $p$ {$<0.001$}} \\
& \multicolumn{7}{l}{\textbf{Male vs Female:} $\beta$ (log-odds): \textbf{-0.176}, OR (odds ratios):  0.84, $p$: {$0.010$}} \\
\midrule


\multirow{6}{*}{\rotatebox{90}{Anger (RAF-DB)}} & \multirow{2}{*}{\rotatebox{90}{Before}}

& Male   &  308 &  146 &  454 &  67.84 &   0.00 \\
& & Female & 142 &   91 &  233 & 60.94 &  -6.90 \\

& \multirow{2}{*}{\rotatebox{90}{After}}

& Male   & 76 & 31 & 107 & 71.02 & 0.00 \\
& & Female & 369 &  211 & 580 & 63.62 & -7.41 \\
& \multicolumn{7}{l}{\textbf{Before vs After:} $\beta_0$ (log-odds): -0.125, OR (odds ratios): 0.88, $p$: {$ 0.314$}} \\
& \multicolumn{7}{l}{\textbf{Male vs Female:} $\beta_1$ (log-odds): 0.314, OR (odds ratios):  1.37, $p$: {$0.018$}} \\


\bottomrule
\end{tabular}
\label{tab:all_captioning}
\vspace{-1em}
\end{table}

In the previous section, we present analyses based on regional activations of activity and emotion categories. On the other hand, in this section, we will analyse the prediction success rates, the match between the model's selected admissible label and the dataset's ground-truth,  with demographic parity and descriptive statistics as presented in Table ~\ref{tab:all_captioning}.

\subsubsection{Activities}
Demographic parity and logistic regression results for activities are shown in Table~\ref {tab:all_captioning}.  Regression analysis for \emph{Sports} in the PHASE dataset indicates significant effects for both the looping stage ($\beta_0 = -0.728$, $p < 0.001$) and gender ($\beta_1 = 0.386$, $p < 0.001$). Images of male subjects have 1.47 times higher odds of successful prediction. The DP gap increases after looping by approximately 4 percentage points, indicating that while overall accuracy improves for both genders after looping (from $\sim$67-71\% to $\sim$76-84\%), the gender disparity worsens: female success rates are lower than those of males.

For the \emph{Caring} activity, regression shows a significant effect of the looping stage ($\beta_0 = -0.914$, $p = 0.005$) but no significant gender effect ($\beta_1 = -0.221$, $p = 0.584$). However, small sample sizes (22 - 37 per group) limit the reliability of these findings.

\subsubsection{Emotions}
Demographic parity and logistic regression results for emotions are shown in Table~\ref{tab:all_captioning}. For \emph{Happiness}, the two datasets show contrasting patterns. For PHASE, looping reduces demographic parity (from +10.96 to +4.88 percentage points). Regression analysis results confirm that both the looping stage ($\beta_0 = 0.286$, $p < 0.001$) and gender ($\beta_1 = -0.541$, $p < 0.001$) significantly affect prediction success, with images of female subjects having higher success rates. In contrast, RAF-DB shows an increase in the DP gap after looping (from +1.94 to +6.53 percentage points). Notably, overall success rates drop dramatically after looping (from $\sim$82 - 85\% to $\sim$47 - 53\%). Regression confirms significant effects for both the looping stage ($\beta_0 = 1.593$, $p < 0.001$) and gender ($\beta_1 = -0.176$, $p = 0.010$), with images of female subjects succeeding more, similar to PHASE.

For \emph{Anger}, although images of male subjects had consistently higher rates for both PHASE and RAF-DB across stages, regression analysis found no significant effects for either looping stage or gender in either dataset. 
\section{Discussion}
\label{sec:discussion}

Across both RAF-DB and PHASE, we find that iterative exchange between a text-to-image generator and an image-to-text model can introduce demographic drift in the generated outputs, 
with this effect being most apparent in age for both action and affect domains. 
Complementing these distributional results, our explainability analysis shows that some concept predictions are supported by spurious evidence (e.g., hair or background rather than action- or expression-relevant cues; see Figure~\ref{fig:emotion_attention_examples_2} and \ref{fig:emotion_attention_examples_4}), providing insights into why looped communication can reinforce demographic drifts.

\subsection{RQ1: Demographic Distribution Drift}
\label{subsec:rq1}

RQ1 asks whether inter-model communication produces systematic \emph{associational} drifts—i.e., whether activities and emotions become disproportionately linked with particular demographic groups through the model's information exchange. Across both benchmarks, we observe consistent demographic drift after the \ac{IGID} loop, with outcomes tending to over-represent younger individuals (for all emotions and activities) and female-presenting individuals (for all emotions); 
while PHASE typically exhibits smaller drifts than RAF-DB. For activities, an interesting observation comes from the Sports category for gender, where the drift was towards generating more female-presenting individuals, as opposed to former stereotypes associating this activity with males~\shortcite{harrison-etal-2023-run,girrbach_large_2025}. We also observe that some activities exhibit more pronounced demographic change than others (e.g., the increase in female representation in \textit{Dance and Music}).  For emotions, the drifts observed in both datasets are consistent with a common gender stereotype that frames women as ``more emotional''~\shortcite{BREBNER2003387}, illustrating how stereotyped associations can surface in inter-model pipelines. From RAF-DB, we also observe increased Black and female-presenting outputs for anger aligned with former stereotypes~\shortcite{jones_aggressive_2016,malveaux_sexual_1979}, yet the same effect was not observed in the Phase dataset, which also shows the model's sensitivity to the initial seed image (RAF-DB images are more close-up) while enforcing these stereotypes. Overall, these results indicate that inter-model interaction can induce distributional drifts that resemble stereotyped associations in some cases, while also producing less intuitive ones that likely reflect on model priors and dataset specifications.



\subsection{RQ2: Regional Associations}
\label{subsec:rq2}


RQ2 asks whether the observed demographic drifts reflect systematic behaviour rather than sampling noise, which we addressed through region-based activation analysis. Overall, activations are often spread across the \textit{face}, \textit{body}, \textit{hair}, and \textit{background}, instead of being consistently concentrated on concept-relevant regions. \textit{Background} information can help in understanding scene context and recognising activities; however, for emotions, facial cues should provide the main evidence. Still, for the \textit{Happiness} emotion, the background contributed the most in both datasets, as shown in Table~\ref{fig:emotion_attention_examples}. Most critically, the \textit{hair} region should be largely irrelevant for both emotions and activities, yet we observe substantial activation in this region for both categories. Activations on \textit{hair} are particularly concerning because this region can correlate with demographic presentation, providing a plausible explanation for the distribution drifts observed through the \ac{IGID} loop. Overall, these patterns suggest that the model may sometimes rely on appearance or contextual associations as ``shortcuts''; in such cases, rather than learning fine-grained emotional cues or activity-related concepts, it may follow biased associations, e.g., inappropriate activations on \textit{hair} for the \emph{Happiness} emotion in Figure~\ref{fig:emotion_attention_examples} may relate to previous stereotypes associating this emotion with females~\cite{hosseini_faces_2025}. 
Taken together, 
when the model grounds predictions in concept-irrelevant cues, demographic-correlated features can influence the inter-model communication and contribute to the observed drifts.

\subsection{RQ3: Prediction Success Rate}
RQ3 asks about the success rates on downstream tasks after the \ac{IGID} loop, which we evaluate using emotion and activity prediction success rates across gender groups. Regression analysis reveals that both the looping stage and gender can significantly affect prediction success, though these effects depend on both the target concept and the dataset setting (Table ~\ref{tab:all_captioning}). For activities, looping improves \emph{Sports} accuracy but worsens gender disparities, consistent with documented stereotypes associating sports with a male-coded activity \shortcite{harrison-etal-2023-run,girrbach_large_2025}, while \emph{Caring} shows no significant gender effects despite being stereotyped as female-coded \shortcite{girrbach_large_2025}.  Accuracy gains for \emph{Sports} may come from the model better recognising stereotypical sports contexts that are themselves male-dominated in existing datasets, and as such, reinforcing rather than correcting existing biases. The finding that accuracy improvements can coincide with fairness degradation, as seen in \emph{Sports}, demonstrates that optimising for performance without explicit fairness constraints does not guarantee equitable outcomes. For emotions, looping reduces gender disparities in PHASE but amplifies them in RAF-DB for \emph{Happiness}, while \emph{Anger} shows no significant effects in either dataset. These divergent patterns suggest that the looping process can amplify or reduce bias depending on what visual features the models learn to rely on. Furthermore, RAF-DB’s close-up portraits may reduce context-related confounds for emotion prediction, whereas PHASE’s broader, context-rich scenes can introduce additional cues that affect model behaviour. Taken together, our findings highlight that after the \ac{IGID} loop, prediction success can differ by gender in concept- and setting-dependent ways, underscoring the need to match dataset conditions to task requirements (e.g., close-up portraits versus wider scenes) and to actively monitor fairness throughout iterative processes.

\section{Future Directions Towards Bias Mitigation}
\label{sec:mitigation}

Taken together, bias in compound systems may not be fully understood by analysing individual components in isolation: when a generator and a recogniser are composed into an interactive loop, demographic drift can emerge, and concept predictions may be supported by spurious visual evidence rather than concept-relevant cues. These observations make bias mitigation particularly pressing for \emph{intermodel} and \emph{compositional} AI systems. 
Building on this, we outline mitigation directions spanning \emph{data-centric}, \emph{training-time}, and \emph{deployment-time/post-training} interventions, with a particular focus on looped model-to-model communication—where, for example, demographic stereotypes or biased associations can be propagated and reinforced across iterations.

\subsection{Data-centric mitigation}

As highlighted by the dataset sensitivity of our findings, data-centric mitigation is critical for inter-model communication pipelines; any bias in the training data can be repeatedly reproduced by the generator model and then reintroduced in subsequent iterations via the recogniser model. 
A direct response to this can be improving coverage of protected attributes across labels through data augmentation \shortcite{lee_survey_2023}; in particular, \emph{counterfactual data augmentation} and \emph{counterfactual data substitution} can create more balanced training variants by swapping demographic indicators while preserving the target concept \shortcite{maudslay_mitigating_2020}. On the other hand, when counterfactuals are insufficient or unrealistic, \emph{targeted data collection} can support underrepresented or anti-stereotypical instances, and can help to mitigate spurious correlations  
\shortcite{dinan_queens_2020,dixon_measuring_2018}. Inspired by this, in our use cases, a practical step is to (i) rebalance each emotion/activity label across demographic groups and diverse contexts through lightweight augmentation or targeted collection, and (ii) maintain a small, balanced evaluation set with demographic-agnostic variants (e.g., blurring hair/face regions or background) to verify that predictions remain stable when demographic cues are removed—reducing stereotype-linked co-occurrences and encouraging the loop to rely on concept-relevant evidence rather than demographic proxies. 

\subsection{Training-time mitigation}

Training-time approaches can help mitigate drift in inter-model communication by avoiding ``shortcuts'' (e.g., using background/hair to predict emotion) and instead forcing the learning of granular, informative nuances. 
Specifically, training-time interventions can reduce demographic sensitivity and discourage spurious correlates in learned representations, e.g., \emph{attribute-aware} training \shortcite{lee_survey_2023,xu_investigating_2020}, which incorporates sensitive attributes (e.g., gender, age, race/skin tone) as variables in group-conditional objectives, can calibrate and monitor the performance across demographic groups. A complementary in-processing strategy is \emph{adversarial training} \shortcite{elazar_adversarial_2018,lee_survey_2023}, in which an auxiliary adversary tries to predict protected attributes from the model’s internal representations while the main model is trained to prevent this, thereby avoiding demographic cues to dominate concept predictions. 
In our setting, this could be used to break loop-reinforced demographic associations, such as \textit{Dance \& Music} and \textit{Happiness} becoming disproportionately linked to female depictions, and to mitigate drift toward younger representations after inter-model exchange. 
Finally, bias can also be addressed at the representation level by using gender-neutral embeddings that isolate or attenuate gender information in learned language representations~\shortcite{zhao2018gender}.

\subsection{Deployment-time and post-training mitigation}

Deployment-time and post-training mitigation are particularly practical for inter-model communication of large generative models without modifying or fine-tuning the backbone models. 
A range of post-training strategies could be used to achieve this: For the generative component, instruction-based methods such as \emph{Fair Diffusion} can provide one route to encourage fairer generation at inference time \shortcite{friedrich_fair_2023}. \emph{Group tagging} or control tokens can offer another route by explicitly conditioning outputs with demographic specifications \shortcite{lee_survey_2023,vanmassenhove_getting_2018}; e.g., in text generation, these tags have been used to control or reduce gendered realisations \shortcite{vanmassenhove_getting_2018} and in our setting, they could be used to request the same emotion or activity across different demographic groups to avoid stereotyped defaults. 
More broadly, \emph{prompt engineering} provides lightweight controls for both components of the loop: persona or user-context conditioning can influence affective predictions \shortcite{lee_analyzing_2024}, while structured prompts (e.g., chain-of-thought) and explicit fairness-oriented instructions can be used 
to guide generation away from demographic drift \shortcite{lee_survey_2023,etesam_contextual_2024,friedrich_fair_2023}. Additionally, explainability can be operationalised as a guardrail~\shortcite{chu2024causal}: if token-conditioned attributions for a target concept consistently concentrate on regions that should be concept-irrelevant (e.g., hair for affect), the system can flag the exchange as unreliable and apply a corrective step such as regeneration with revised prompts, stronger constraints, or a different model pairing. Similarly, drift metrics can be monitored continuously to trigger intervention when demographic distributions deviate beyond a threshold. Overall, these interface-level strategies can serve as practical guardrails through the knowledge transfer between interconnected models that reduce the propagation of spurious or demographically correlated cues across iterations.

\section{Conclusion and Considerations}

In this paper, we investigate how associational bias can drift over repeated model-to-model exchanges, focusing on human affect and activities. We introduce two complementary pipelines: (i) an inter-model communication pipeline that alternates between text-to-image generation and image-to-text description to quantify demographic distribution shifts in loop outputs, and (ii) an explainability pipeline that uses token-conditioned saliency and region-based aggregation to assess whether predictions are grounded in concept-relevant evidence or in spurious visual cues. Across RAF-DB and PHASE datasets for human activity and emotions, we observe consistent demographic drifts in loop-generated images—often trending toward younger, female-presenting representations—with variation by dataset and category. The explainability results show that concept predictions are not always grounded in semantically appropriate evidence (e.g., emotion cues on the face), and can instead rely on background or hair cues, which can help explain why the loop outputs drift toward certain demographic groups.

This work raises ethical and practical considerations for real-world deployment. First, the demographic analyses in Section~\ref{sec:results} rely on visual assumptions of gender, age, and ethnicity/skin tone; they should be interpreted as perceived attributes rather than ground truth. 
Second, an important consideration is that our results—alongside prior work reporting demographic biases in widely used foundation models \shortcite{esser_scaling_2024}—suggest that the ``plug-and-play'' use of such models in activity recognition or affective computing pipelines may 
affect the system's fairness in downstream tasks. Accordingly, deployments should include regular bias checks across demographic groups, 
and while doing this, former findings emphasise the importance of transparency and explainability to make such biases easier to detect, diagnose, and address~\shortcite{van2023deep}. More broadly, addressing these issues will likely require not only technical interventions but also clearer governance and responsible-use guidance to support the ethical integration of generative AI systems \shortcite{hernandez_guidelines_2021}.



\begin{acks}
The authors have applied a Creative Commons Attribution (CC BY) licence to any Author Accepted Manuscript version arising. The authors also thank Rahma Elsheikh for the fruitful discussions.
\noindent\textbf{Data access:} 
This study involved secondary analyses of existing datasets. All datasets are described and cited accordingly. 
\noindent\textbf{Funding:} The work of F. I. Dogan \& H. Gunes were supported in part by CHANSE and NORFACE through the MICRO project, funded by ESRC/UKRI (grant ref. UKRI572).
\noindent\textbf{Contributions:} Conceptualisation: FID, HG. Methodology: FID, YW, KP. Data curation: YW.  Investigation: FID, YW, KP. Software: YW, KP. Formal analysis \& Visualisation: FID, YW, KP. Resources: FID, JC, HG. Writing – original draft: FID, YW, KP, JC. Writing – review \& editing: FID, JC, HG. Supervision: FID, HG. Project administration: FID, HG. Funding acquisition: HG.
\end{acks}

\printbibliography

\newpage
\appendix
\section*{APPENDIX}
\section{Text-seeded run results}
\label{app:text_seeded_results}

\begin{figure*}[h!]
    \centering
    \includegraphics[width=\linewidth]{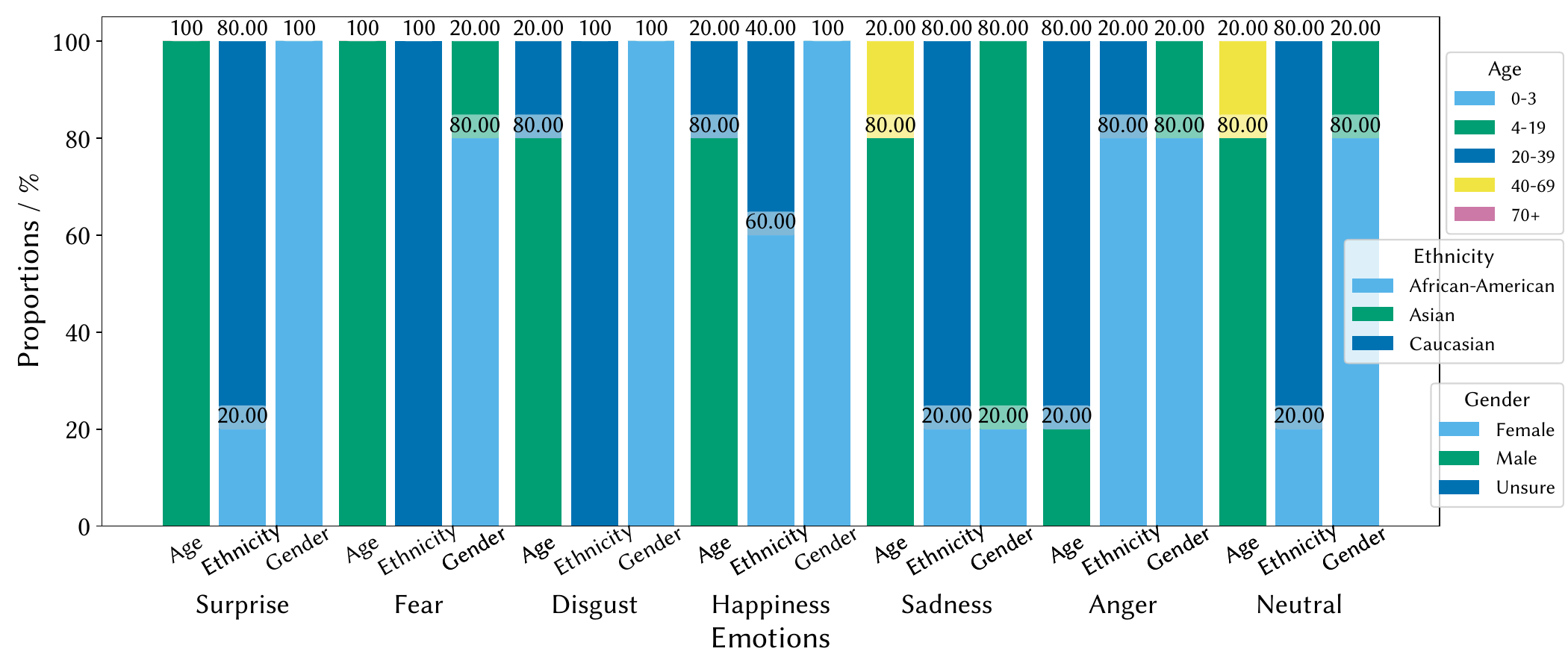}
    \vspace{-.5cm}
    \caption{Gender, ethnicity, and age data for five iterations of a text-seeded \ac{IGID} loop per emotion.}
    \label{fig:5_iters}
\end{figure*}

In addition to the main image-seeded findings reported in the manuscript, we also run small-scale text-seeded experiments to further evaluate the \ac{IGID} loop under an alternative initialisation. As shown in Figure~\ref{fig:5_iters}, the text-seeded runs exhibit a clear qualitative skew in the generated people: across emotions, outputs tend to be predominantly female-presenting, relatively young, and disproportionately lighter-skinned/white, with only limited counter-examples that vary by emotion category.


We also observe a failure mode for \textit{Sadness}, where several iterations yield visually ambiguous subjects (e.g., silhouette-like renderings), making demographic annotation unreliable for those samples.

Because this study is small in scale (five iterations per emotion), it is not intended for statistical inference. Nevertheless, the direction of these qualitative patterns is consistent with the demographic drifts observed in our large-scale RAF-DB and PHASE experiments. We therefore report the text-seeded results in the appendix as supporting qualitative evidence rather than as a primary result.

\section{Looping and Convergence}
\label{subsec:exploratory_analysis}

\begin{figure*}[b!]
\begin{subfigure}{.48\textwidth}
  \centering
  \includegraphics[width=\linewidth]{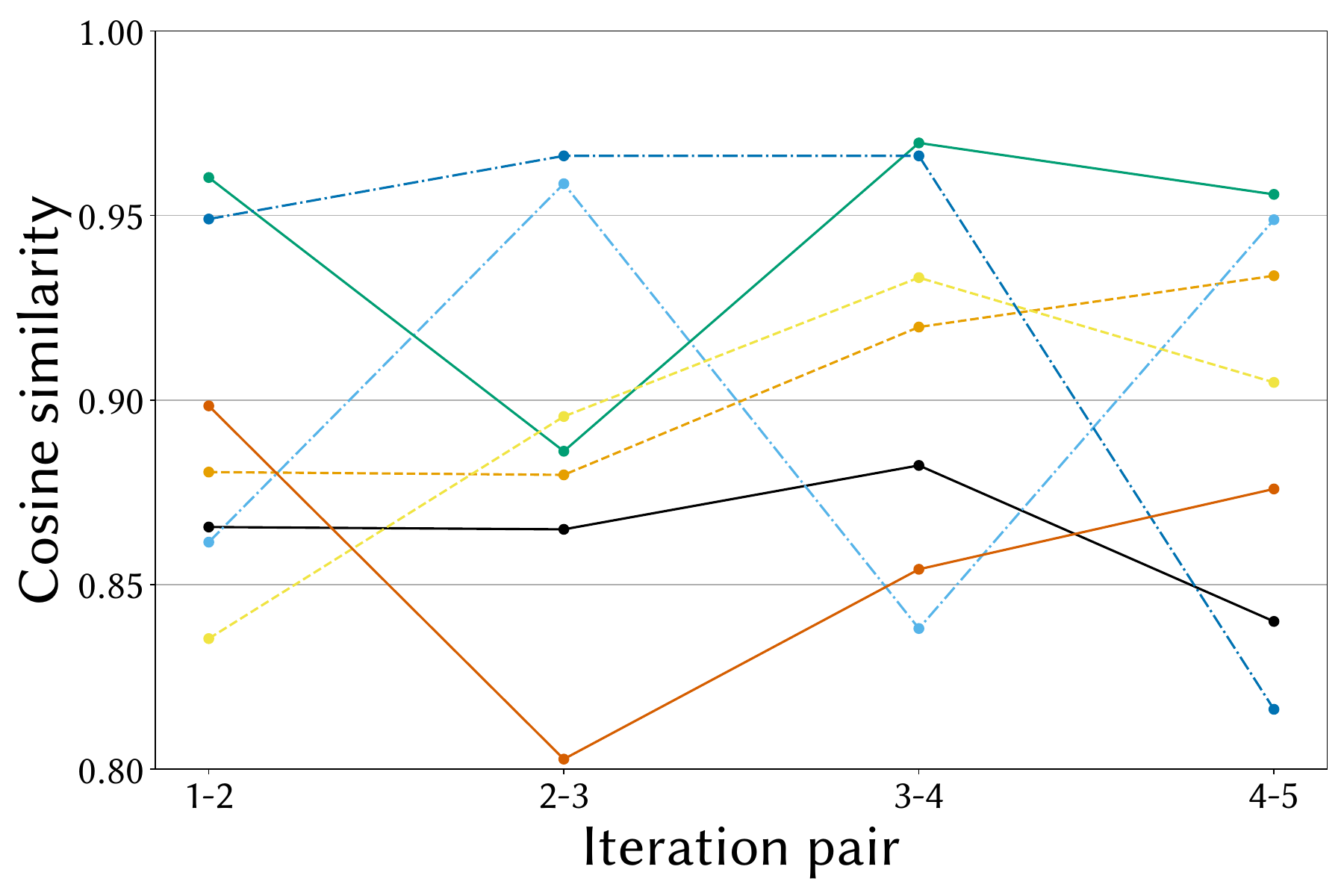}
  \caption{Pairwise text similarities (text-seeded)}
  \label{fig:sub_txt_desc_desc}
\end{subfigure}%
\begin{subfigure}{.48\textwidth}
  \centering
  \includegraphics[width=\linewidth]{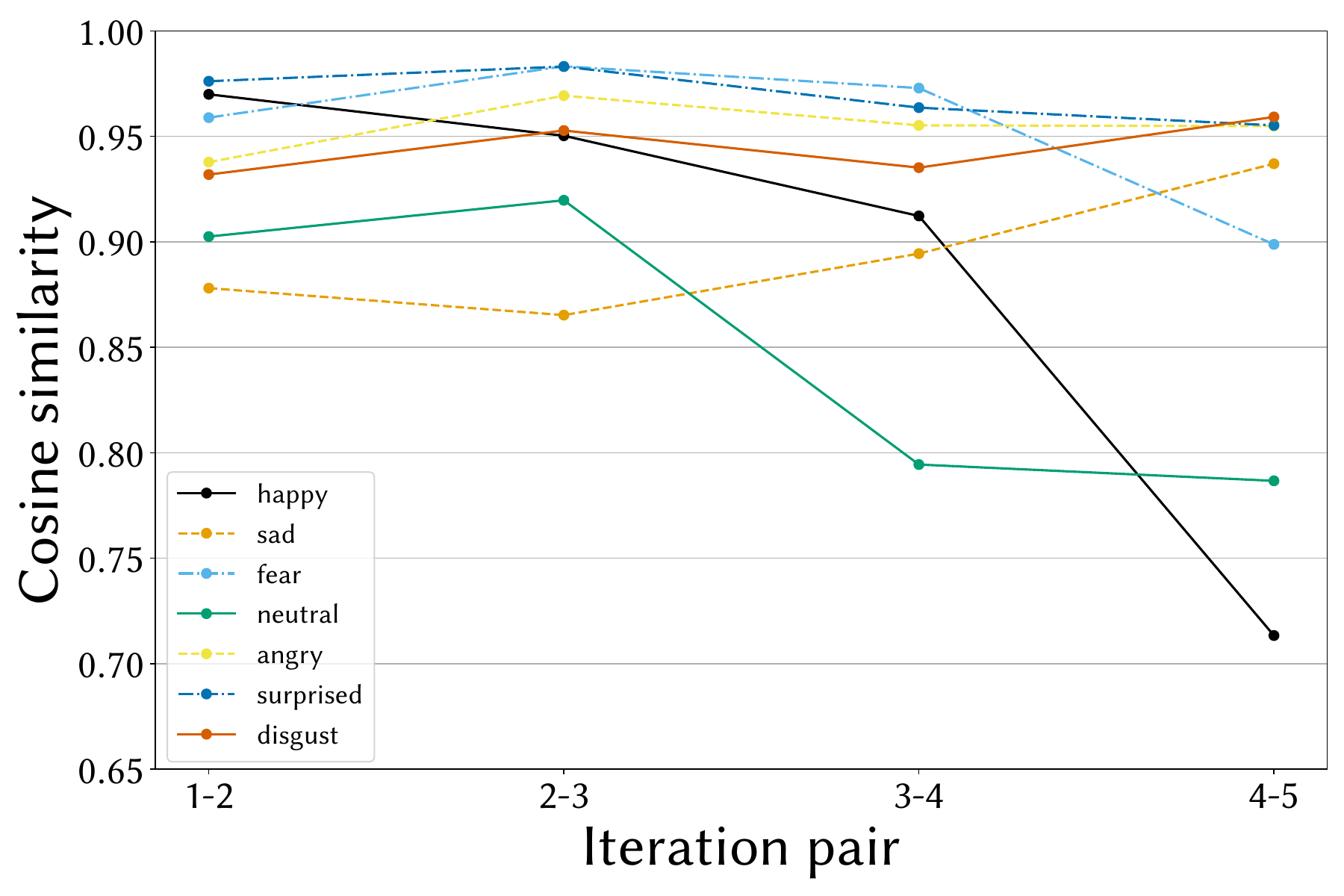}
  \caption{Pairwise image similarities (text-seeded)}
  \label{fig:sub_txt_img_img}
\end{subfigure}

\vspace*{0.5cm}

\begin{subfigure}{.48\textwidth}
  \centering
  \includegraphics[width=\linewidth]{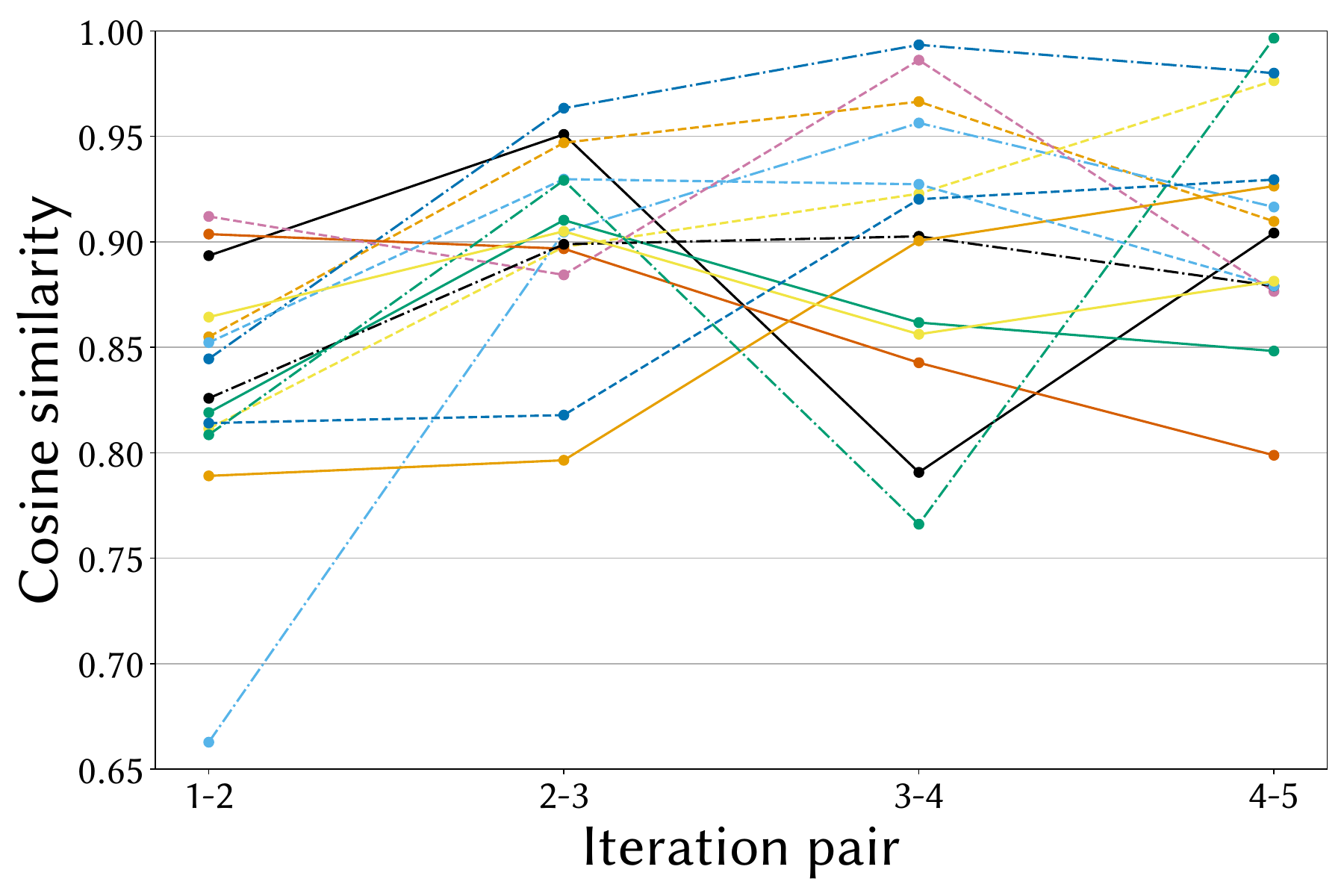}
  \caption{Pairwise text similarities (RAF-seeded)}
  \label{fig:sub_raf_desc_desc}
\end{subfigure}%
\begin{subfigure}{.48\textwidth}
  \centering
  \includegraphics[width=\linewidth]{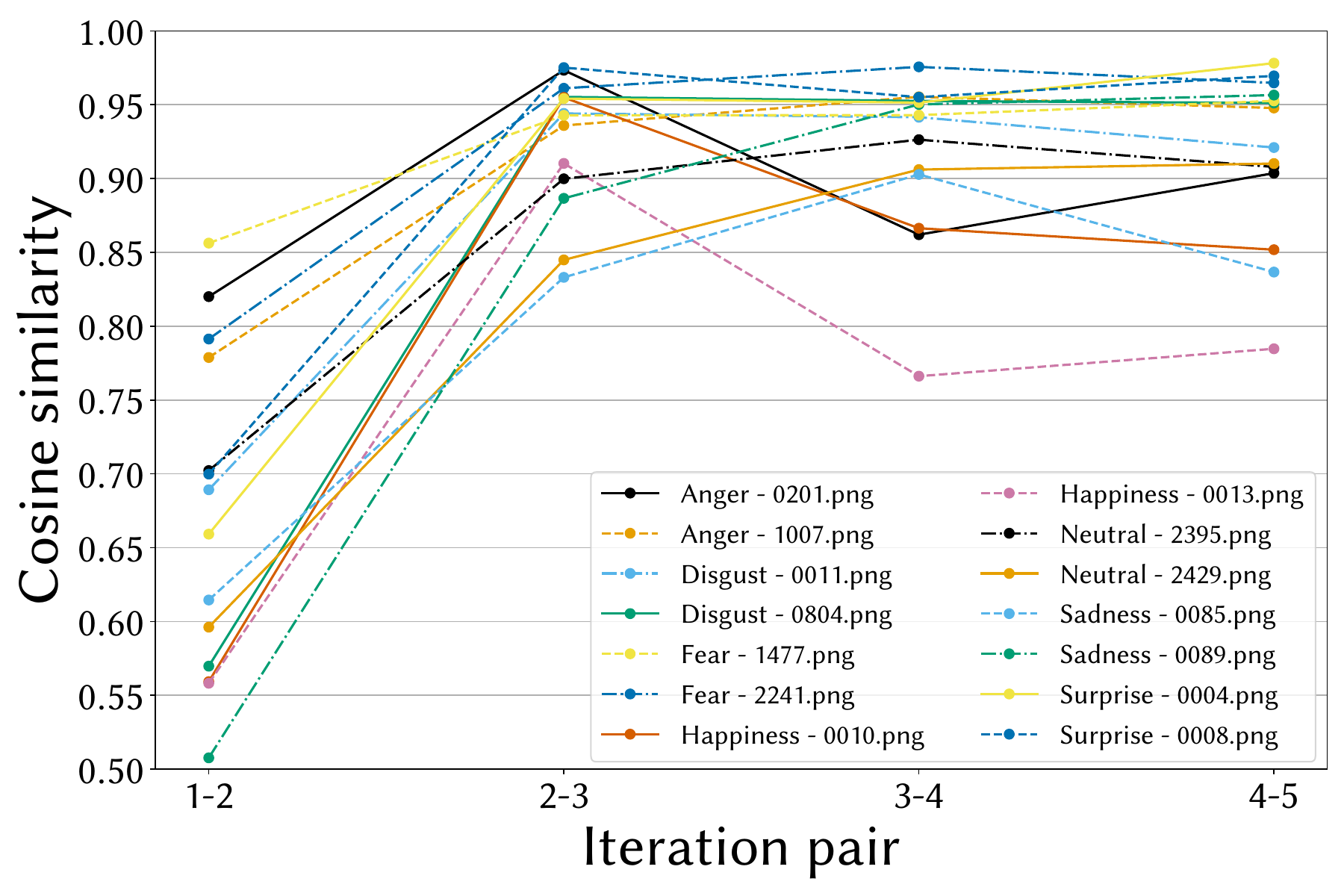}
  \caption{Pairwise image similarities (RAF-seeded)}
  \label{fig:sub_raf_img_img}
\end{subfigure}
\caption{Pairwise text-to-text (\subref{fig:sub_txt_desc_desc}) and image-to-image (\subref{fig:sub_txt_img_img}) cosine similarities for an \ac{IGID} loop initialised with a text prompt. (\subref{fig:sub_raf_desc_desc}) and (\subref{fig:sub_txt_img_img}) show equivalent statistics for an \ac{IGID} loop initialised with RAF-DB images.}
\label{fig:txt_pairwise_sim}
\end{figure*}

This section quantifies how quickly the \ac{IGID} loop stabilises by measuring similarity between consecutive text and image outputs across iterations, for both text-seeded initialisation and image-seeded initialisation (where the loop is initialised from randomly selected RAF-DB images for different emotions).

We assess convergence of the \ac{IGID} mechanism by computing pairwise cosine similarities between consecutive iterations. For the text stream, we embed each generated description using a sentence-embedding model\footnote{Model: \url{https://huggingface.co/sentence-transformers/all-MiniLM-L6-v2}.} and measure cosine similarity between $d_t$ and $d_{t-1}$. For the image stream, we embed each generated image using CLIP \shortcite{radford_learning_2021} and measure cosine similarity between $\mathit{im}_t$ and $\mathit{im}_{t-1}$. This provides a quantitative view of how outputs evolve as the loop iterates.

Figure~\ref{fig:txt_pairwise_sim} summarises these pairwise similarities. The left-hand plots (Figures~\ref{fig:sub_txt_desc_desc} and \ref{fig:sub_raf_desc_desc}) report description-to-description similarities, which remain consistently high across emotions (typically above 0.8), indicating that the generated text stabilises quickly across iterations. The right-hand plots (Figures~\ref{fig:sub_txt_img_img} and \ref{fig:sub_raf_img_img}) show image-to-image similarities; despite greater variability in the visual domain, similarities generally rise above 0.75 from the second iteration onward (with \textit{Happy} in the text-seeded setting as a notable exception). Overall, consecutive iterations exhibit strong overlap in both modalities, suggesting rapid convergence of the loop.

Motivated by this convergence behaviour, we run a single \ac{IGID} loop for the large-scale RAF-DB and PHASE experiments in the main paper. We retain five iterations for the smaller text-seeded study to characterise iterative dynamics under text initialisation.

\section{Significance Tests on Demographic Distribution Drifts}
\label{significance_appendix}


This section reports the significance-testing results underlying the demographic distribution drift analyses in the main paper, which address \textit{RQ1} (Section~\ref{findings-RQ1}). For each dataset and domain, we apply the Stuart--Maxwell test of marginal homogeneity to compare demographic label distributions before versus after inter-model exchange, and we report the number of samples ($N$), the raw p-values ($p$), and Benjamini--Hochberg adjusted p-values ($q$) with $\alpha = 0.01$ to control the false discovery rate across multiple comparisons. Tables~\ref{tab:phase-act-pvals}--\ref{tab:raf-pvals} provide these values for PHASE activities, PHASE emotions, and RAF-DB emotions, respectively.

{
\begin{table*}[b!]
\centering
\scriptsize
\sisetup{
  scientific-notation = true,
}
\caption{Number of samples (N), raw p-values ($p$), and Benjamini-Hochberg adjusted p-values ($q$, with $\alpha = 0.01$) for PHASE Activities Stuart-Maxwell homogeneity scores.}
\label{tab:phase-act-pvals}
\begin{tabular}{
l
l
S[table-format=1.3e2]
S[table-format=1.3e2]
l
S[table-format=1.3e2]
S[table-format=1.3e2]
l
S[table-format=1.3e2]
S[table-format=1.3e2]
}
\toprule
& \multicolumn{3}{c}{Age} & \multicolumn{3}{c}{Ethnicity} & \multicolumn{3}{c}{Gender} \\
\cmidrule(lr){2-4}
\cmidrule(lr){5-7}
\cmidrule(lr){8-10}
Activity
& {$N$}
& {$p$}
& {$q$}
& {$N$}
& {$p$}
& {$q$}
& {$N$}
& {$p$}
& {$q$} \\
\midrule
Helping \& Caring & 60 & \num{1.2e-01} & \num{1.8e-01} & 58 & \num{3.2e-01} & \num{3.8e-01} & 63 & \num{2.8e-01} & \num{3.6e-01} \\
Eating & 150 & \num{1.2e-04} & \num{2.2e-04} & 148 & \num{5.5e-01} & \num{6.3e-01} & 153 & \num{3.3e-02} & \num{5.1e-02} \\
Household & 9 & \num{1.0e+00} & \num{1.0e+00} & 9 & \num{1.0e+00} & \num{1.0e+00} & 9 & \num{1.0e+00} & \num{1.0e+00} \\
Dance \& Music & 3044 & {$<$} \num{1e-18} & {$<$} \num{1e-18} & 3047 & \num{1.1e-12} & \num{3.1e-12} & 3085 & {$<$} \num{1e-18} & {$<$} \num{1e-18} \\
Personal Care & 23 & \num{2.2e-01} & \num{3.2e-01} & 23 & \num{3.2e-01} & \num{3.8e-01} & 23 & \num{6.5e-01} & \num{7.3e-01} \\
Posing & 5846 & {$<$} \num{1e-18} & {$<$} \num{1e-18} & 5832 & \num{1.3e-10} & \num{3.3e-10} & 5850 & {$<$} \num{1e-18} & {$<$} \num{1e-18} \\
Sports & 3336 & {$<$} \num{1e-18} & {$<$} \num{1e-18} & 3482 & \num{6.2e-04} & \num{1.1e-03} & 3553 & {$<$} \num{1e-18} & {$<$} \num{1e-18} \\
Transportation & 122 & \num{4.1e-03} & \num{6.8e-03} & 112 & \num{2.5e-01} & \num{3.4e-01} & 130 & \num{2.3e-05} & \num{4.7e-05} \\
Work & 587 & \num{1.4e-08} & \num{3.0e-08} & 589 & \num{3.1e-09} & \num{7.1e-09} & 633 & \num{1.1e-16} & \num{3.7e-16} \\
\bottomrule
\end{tabular}
\end{table*}
}

{
\begin{table*}[b!]
\centering
\scriptsize
\sisetup{
  scientific-notation = true,
}
\caption{Number of samples (N), raw p-values ($p$), and Benjamini-Hochberg adjusted p-values ($q$, with $\alpha = 0.01$) for PHASE Emotions Stuart-Maxwell homogeneity scores.}
\label{tab:phase-ems-pvals}
\begin{tabular}{
l
l
S[table-format=1.3e2]
S[table-format=1.3e2]
l
S[table-format=1.3e2]
S[table-format=1.3e2]
l
S[table-format=1.3e2]
S[table-format=1.3e2]
}
\toprule
& \multicolumn{3}{c}{Age} & \multicolumn{3}{c}{Ethnicity} & \multicolumn{3}{c}{Gender} \\
\cmidrule(lr){2-4}
\cmidrule(lr){5-7}
\cmidrule(lr){8-10}
Emotion
& {$N$}
& {$p$}
& {$q$}
& {$N$}
& {$p$}
& {$q$}
& {$N$}
& {$p$}
& {$q$} \\
\midrule
Anger & 238 & {$<$} \num{1e-18} & {$<$} \num{1e-18} & 237 & \num{3.1e-01} & \num{3.3e-01} & 241 & {$<$} \num{1e-18} & {$<$} \num{1e-18} \\
Fear & 68 & \num{1.4e-03} & \num{2.1e-03} & 67 & \num{5.1e-01} & \num{5.1e-01} & 68 & \num{1.4e-02} & \num{1.8e-02} \\
Happy & 6857 & {$<$} \num{1e-18} & {$<$} \num{1e-18} & 6846 & \num{1.2e-02} & \num{1.6e-02} & 6870 & {$<$} \num{1e-18} & {$<$} \num{1e-18} \\
Neutral & 8513 & {$<$} \num{1e-18} & {$<$} \num{1e-18} & 8564 & {$<$} \num{1e-18} & {$<$} \num{1e-18} & 8678 & {$<$} \num{1e-18} & {$<$} \num{1e-18} \\
Sad & 224 & \num{2.3e-07} & \num{4.3e-07} & 221 & \num{9.0e-02} & \num{1.0e-01} & 222 & \num{3.4e-07} & \num{5.7e-07} \\
\bottomrule
\end{tabular}
\end{table*}
}

{
\begin{table*}[b!]
\scriptsize
\centering
\sisetup{
  scientific-notation = true,
}
\caption{Number of samples (N), raw p-values ($p$), and Benjamini-Hochberg adjusted p-values ($q$, with $\alpha = 0.01$) for RAF-DB Emotions Stuart-Maxwell homogeneity scores.}
\label{tab:raf-pvals}
\begin{tabular}{
l
l
S[table-format=1.3e2]
S[table-format=1.3e2]
l
S[table-format=1.3e2]
S[table-format=1.3e2]
l
S[table-format=1.3e2]
S[table-format=1.3e2]
}
\toprule
& \multicolumn{3}{c}{Age} & \multicolumn{3}{c}{Ethnicity} & \multicolumn{3}{c}{Gender} \\
\cmidrule(lr){2-4}
\cmidrule(lr){5-7}
\cmidrule(lr){8-10}
Emotion
& {$N$}
& {$p$}
& {$q$}
& {$N$}
& {$p$}
& {$q$}
& {$N$}
& {$p$}
& {$q$} \\
\midrule
Surprise & 1619 & {$<$} \num{1e-18} & {$<$} \num{1e-18} & 1619 & {$<$} \num{1e-18} & {$<$} \num{1e-18} & 1619 & {$<$} \num{1e-18} & {$<$} \num{1e-18} \\
Fear & 355 & {$<$} \num{1e-18} & {$<$} \num{1e-18} & 355 & \num{7.3e-10} & \num{7.3e-10} & 355 & {$<$} \num{1e-18} & {$<$} \num{1e-18} \\
Disgust & 877 & {$<$} \num{1e-18} & {$<$} \num{1e-18} & 877 & {$<$} \num{1e-18} & {$<$} \num{1e-18} & 877 & {$<$} \num{1e-18} & {$<$} \num{1e-18} \\
Happiness & 5954 & {$<$} \num{1e-18} & {$<$} \num{1e-18} & 5956 & {$<$} \num{1e-18} & {$<$} \num{1e-18} & 5956 & {$<$} \num{1e-18} & {$<$} \num{1e-18} \\
Sadness & 2459 & {$<$} \num{1e-18} & {$<$} \num{1e-18} & 2459 & {$<$} \num{1e-18} & {$<$} \num{1e-18} & 2459 & {$<$} \num{1e-18} & {$<$} \num{1e-18} \\
Anger & 867 & {$<$} \num{1e-18} & {$<$} \num{1e-18} & 867 & \num{5.6e-15} & \num{5.8e-15} & 867 & {$<$} \num{1e-18} & {$<$} \num{1e-18} \\
Neutral & 3203 & {$<$} \num{1e-18} & {$<$} \num{1e-18} & 3203 & {$<$} \num{1e-18} & {$<$} \num{1e-18} & 3203 & {$<$} \num{1e-18} & {$<$} \num{1e-18} \\
\bottomrule
\end{tabular}
\end{table*}
}

\end{document}